\documentclass[letterpaper,twocolumn,aps,notitlepage,longbibliography,superscriptaddress,footinbib,reprint,prb]{revtex4-2}

\usepackage{xcolor}
\usepackage{etoolbox,environ,listofitems}
\usepackage{amsmath,amsfonts,amssymb,amsthm,bm,bbm,stmaryrd}
\usepackage{graphicx,tikz,pgfplots,tikz-3dplot}
\usepackage{environ,xstring}
\usepackage{apptools,pdftexcmds}
\usepackage{tabularx,multirow}
\usepackage{booktabs,colortbl}
\usepackage{adjustbox,soul,diagbox,cancel,hhline}
\usepackage{outlines,changepage,ftnxtra}
\usepackage[letterpaper,centering,hmargin=2cm,vmargin=2cm]{geometry}

\usepackage[font=small,textfont=normalfont,singlelinecheck=off]{caption}
\usepackage[font=small,textfont=normalfont,singlelinecheck=off,justification=raggedright,labelfont=bf]{subcaption}

\usepackage[pdfpagelabels,pdftex,bookmarks,breaklinks]{hyperref}

\usepackage[capitalise]{cleveref}
\usepackage[]{footmisc}

\makeatletter

\patchcmd{\@hex@@Hex}{f\else}{F\else}{\typeout{Patching xcolor}}{}

\def\orcid#1{
	\ifx&#1&%
	\else
	\@AF@join{
		\twocolumn@sw{\newline}{\kern-.25em}
		\href{https://orcid.org/#1}{ORCID:#1}}%
	\fi
}%

\def\@homepage#1#2{%
	\endgroup
	\ifx&#2&%
	\else
	\@AF@join{#1\href{https://#2}{#2}}%
	\fi
}%

\let\@@cite\cite
\renewcommand\cite[1]{\ifstrempty{#1}{\@@cite{blank}}{\@@cite{#1}}}
\let\oldonlinecite\onlinecite

\renewcommand{\onlinecite}[1]{%
\readlist*{\@citelist}{#1}%
\ifnumgreater{\@citelistlen}{1}{Refs}{Ref}.~[\oldonlinecite{#1}]%
}

\AtAppendix{
	\let\oldsection\section
	\renewcommand{\section}[1]{
		\clearpage
		\oldsection{#1}
	}
	\@addtoreset{figure}{section}
	
}

\newcounter{subsubsubsection}[subsubsection]
\def\subsubsubsectionmark#1{}

\def\@subsubsubsection{\@startsection{subsubsubsection}{4}{5pt}{3ex}{2ex}{\small\centering\emph}}
\def\l@subsubsubsection{\@dottedtocline{4}{4.8em}{4.2em}}

\def\subsubsubsection{\@ifstar{\@subsubsubsection*}{\@subsubsubsection*}}

\let\@oldtitle\title
\def\title#1{
	\def\@pdftitle{#1}
	\@oldtitle{#1}
}

\let\@oldauthor\author
\def\author#1{
	\ifx\@pdfauthor\empty
		\def\@pdfauthor{#1}
	\else
		\expandafter\def\expandafter\@pdfauthor\expandafter{\@pdfauthor, #1}
	\fi
	\@oldauthor{#1}
}
\makeatother

\definecolor{layer1}{RGB}{100, 0, 0}
\definecolor{layer2}{RGB}{0, 100, 0}
\definecolor{layer3}{RGB}{0, 0, 100}

\definecolor{yellow}{RGB}{204, 204, 0}
\definecolor{darkblue}{RGB}{0,0,127}
\definecolor{darkgreen}{RGB}{0,180,0}
\definecolor{darkred}{RGB}{180,0,0}
\definecolor{nicegreen}{RGB}{5, 173, 39}
\definecolor{nightBlue}{RGB}{27,81,166}

\hypersetup{
	colorlinks,
	linkcolor=nightBlue,
	citecolor=nightBlue,
	filecolor=red,
	urlcolor=nightBlue,
	pdftitle={\@pdftitle},
	pdfauthor={\@pdfauthor},
	pdfcreator={LaTeX with extended hyperref}
}

\crefname{enumi}{step}{steps}
\Crefname{enumi}{Step}{Steps}

\theoremstyle{plain}
\newtheorem{theorem}{Theorem}

\theoremstyle{definition}
\newtheorem{definition}[theorem]{Definition}

\newcommand{\Stab}{\mathcal{S}}
\renewcommand{\S}{\Stab}

\newcommand{\bareLog}{\mathcal{L}_{\text{bare}}}
\newcommand{\dressedLog}{\mathcal{L}_{\text{dressed}}}

\newcommand{\bareDistance}[1][]{D_{\text{bare}}^{#1}}
\newcommand{\dressedDistance}[1][]{D_{\text{dr}}^{#1}}

\newcommand{\Pauli}[2]{
	\mathcal{P}_{\!#1}^{#2}
}

\newcommand{\Torus}[1]{
	\operatorname{\bf T}^{#1}
}

\newcommand{\I}{\ensuremath{[0,1]}}

\newcommand{\Interval}{
	\operatorname{\bf I}
}

\newcommand{\GCC}{gauge color code}
\newcommand{\STC}{subsystem toric code}
\newcommand{\SAQD}{SAQD}

\DeclareMathOperator{\rk}{rk}
\DeclareMathOperator{\Span}{span}

\newcommand{\iso}{\simeq}

\newcommand{\ZZ}[1]{\mathbb{Z}_{#1}}

\newcommand{\ket}[1]{|{#1}\rangle}

\newcommand{\D}[1]{\mathcal{D}{} }

\newcommand{\comm}[1]{\left[#1\right]}
\newcommand{\restrict}[1]{\raise-.2ex\hbox{\ensuremath|}_{#1}}

\newcommand{\G}{\mathcal{G}}

\usepackage[mathscr]{eucal}
\newcommand{\graph}{\mathscr{G}}

\newcommand{\vvec}[2][]
{\ensuremath{%
		\operatorname{\bf Vec}^{#1}
		\ifstrequal{#2}{}
		{}
		{\left(#2\right)}%
	}}

\newcommand{\rrep}[2][]
{\ensuremath{%
		\operatorname{\bf Rep}^{#1}
		\ifstrequal{#2}{}
		{}
		{\left(#2\right)}%
	}}

\let\originalleft\left
\let\originalright\right
\renewcommand{\left}{\mathopen{}\mathclose\bgroup\originalleft}
\renewcommand{\right}{\aftergroup\egroup\originalright}

\newcommand{\cat}[1]{\ensuremath{\mathcal{#1}}}
\newcommand{\C}{\cat{C}}

\newcommand{\set}[2]{
	\left\{#1
	\ifthenelse{\equal{#2}{}}{}{\,\middle|\,#2}
	\right\}
}

\newcommand{\order}[1]
{
	\left|
	\ifstrequal{#1}{}
	{\bullet}
	{#1}
	\right|
}

\newcommand{\subgp}{\leqslant}

\newcommand{\cent}[1]
{\ensuremath{
		\operatorname{\bf Z}
		\ifstrequal{#1}{}
		{}
		{\left(#1\right)}
	}}

\newcommand{\centralizer}[2]
{\ensuremath{
		\operatorname{\bf C}
		\ifstrequal{#2}{}{_{\Pauli{q}{n}}}{_{#2}}
		\ifstrequal{#1}{}
		{}
		{\left(#1\right)}
	}}

\newcommand{\normalizer}[2]
{\ensuremath{
		\operatorname{\bf N}
		\ifstrequal{#2}{}{_{\Pauli{q}{n}}}{_{#2}}
		\ifstrequal{#1}{}
		{}
		{\left(#1\right)}
	}}

\newcommand{\grouppresentation}[2]{
	\left\langle#1
	\ifthenelse{\equal{#2}{}}{}{\,\middle|\,#2}
	\right\rangle
}

\newcommand{\codeparameters}[2]{
	\left\llbracket#1\right\rrbracket_{#2}
}

\newenvironment{subalign}[1][]{\subequations\label{#1}\align}{\endalign\endsubequations}

\newcommand{\define}[1]{\emph{#1}}

\newcolumntype{C}[1]{>{\centering\let\newline\\\arraybackslash\hspace{0pt}}m{#1}}

\newcommand{\includegraphicsarray}[1]{
	\ensuremath{%
		\begin{array}{c}%
			\includegraphics{#1}%
		\end{array}%
	}
}

\allowdisplaybreaks

\graphicspath{{./figures/}}

\newcommand{\interactivefig}{\href{https://mikevasmer.github.io/qudit-single-shot/}{An interactive version of this figure is available}~\protect\cite{website}.}

\allowdisplaybreaks

\begin{document}
\title{Lifting topological codes: Three-dimensional subsystem codes from two-dimensional anyon models}

\author{Jacob C.\ Bridgeman}
\email{jcbridgeman1@gmail.com}
\homepage{jcbridgeman.github.io}
\orcid{0000-0002-5638-6681}
\affiliation{Perimeter Institute for Theoretical Physics, Waterloo, ON N2L 2Y5, Canada}
\affiliation{Department of Physics and Astronomy, Ghent University, Krijgslaan 281, S9, B-9000 Ghent, Belgium}

\author{Aleksander Kubica}
\email{akubica@caltech.edu}
\homepage{}
\orcid{0000-0001-8213-8190}
\affiliation{AWS Center for Quantum Computing, Pasadena, CA 91125, USA}
\affiliation{California Institute of Technology, Pasadena, CA 91125, USA}

\author{Michael Vasmer}
\email{mikevasmer@gmail.com}
\homepage{mikevasmer.github.io}
\orcid{0000-0002-6711-5924}
\affiliation{Perimeter Institute for Theoretical Physics, Waterloo,  ON N2L 2Y5, Canada}
\affiliation{Institute for Quantum Computing, University of Waterloo, Waterloo, ON N2L 3G1, Canada}
\date{\today}
\defcitealias{Kubica2022}{Kubica and Vasmer, Nat. Commun. \textbf{13}, 6272 (2022)}
\begin{abstract}
	\nocite{website}
	Topological subsystem codes in three spatial dimensions allow for quantum error correction with no time overhead, even in the presence of measurement noise.
	The physical origins of this single-shot property remain elusive, in part due to the scarcity of known models.
	To address this challenge, we provide a systematic construction of a class of topological subsystem codes in three dimensions built from abelian quantum double models in one fewer dimension.
	Our construction not only generalizes the recently introduced subsystem toric code~[\citetalias{Kubica2022}] but also provides a new perspective on several aspects of the original model, including the origin of the Gauss law for gauge flux, and boundary conditions for the code family.
	We then numerically study the performance of the first few codes in this class against phenomenological noise to verify their single-shot property.
	Lastly, we discuss Hamiltonians naturally associated with these codes, and argue that they may be gapless.
\end{abstract}
\maketitle


\section{Introduction}\label{sec:introduction}

In order to reliably store and process quantum information, one needs to protect it against the deleterious effects of noise.
Ideally, one would like to construct a self-correcting quantum memory~\cite{Dennis2002,Brown2016,Terhal2015}, which is a quantum many-body system capable of passively protecting encoded information, analogously to magnetic storage for classical information.
Such protection can be achieved by carefully engineering the Hamiltonian describing the system and the coupling of the system to a thermal bath~\cite{Alicki2010,Chesi2010,Pastawski2011}.
Unfortunately, it is not known how to realize a self-correcting quantum memory in three or fewer spatial dimensions without making strong assumptions about symmetries of the system and noise~\cite{Roberts2020,Kubica2018,Stahl2021}.

A typical solution to the problem of protecting quantum information invokes techniques of quantum error correction (QEC)~\cite{Shor1995,Steane1996}, whereby one encodes information into some QEC code.
By measuring certain operators, commonly referred to as parity checks, one can then detect and, subsequently, correct errors afflicting the encoded information.
Topological codes constitute a particularly appealing class of QEC codes, which can be realized by arranging qubits on some lattice and measuring parity checks that are geometrically local.
Most prominent examples of topological codes include the toric code~\cite{Kitaev2003,Dennis2002} and color code~\cite{Bombin2006}, which can be readily implemented in two spatial dimensions with currently pursued quantum hardware, such as superconducting circuits~\cite{Devoret2013,Blais2021} or Rydberg atoms~\cite{Saffman2010,Browaeys2020}.

In the presence of unavoidable measurement errors, performing reliable QEC becomes more intricate and resource-intensive.
With topological codes in two or three spatial dimensions, one usually resorts to performing repeated rounds of measurements in order to gain confidence in their outcomes~\cite{Dennis2002,Fowler2012}.
In particular, the number of measurement rounds needs to scale with the system size, unless one is willing to tackle a serious engineering challenge of measuring non-local operators of high weight~\cite{Campbell2019,Ashikhmin2020,Delfosse2022}.
As a result, the time overhead of QEC increases and QEC itself becomes susceptible to time-correlated noise.

One can reduce the time overhead and achieve resilience to time-correlated noise by considering unorthodox topological codes, such as the recently-introduced three-dimensional subsystem toric code~\cite{Kubica2022} and gauge color code~\cite{Bombin2015,Kubica2015a,Kubicathesis},
that allow for single-shot QEC~\cite{Bombin2015a}.
Roughly speaking, the single-shot QEC property asserts that, even in the presence of measurement errors, one can perform reliable QEC with a topological code by only performing a constant number of measurement rounds of geometrically-local operators.
However, the origins of the single-shot QEC property remain elusive, with only few topological codes demonstrating it.

In this article, we provide a systematic construction of a novel class of QEC codes, which we refer to as \define{subsystem abelian quantum double} (\SAQD{}) codes.
This class can be viewed as a generalization of the subsystem toric code and is also closely related to the gauge color code.
Roughly speaking, our construction allows to take any abelian quantum double model (that is natively defined in two spatial dimensions) and realize a corresponding topological subsystem code in three spatial dimensions.
In addition, we investigate topological subsystem codes from the perspective of reliable quantum memories capable of single-shot QEC.
In particular, we describe computationally-efficient decoding algorithms and benchmark their performance against phenomenological noise.
We report memory thresholds around $1\%$, which are competitive with that of the surface code~\cite{Wang2003,Ohno2004} given the advantage of single-shot QEC.
Lastly, we argue that the Hamiltonians that naturally arise from topological subsystem codes are probably at a phase transition, providing an indication that these models will not serve as a self-correcting quantum memory.

We remark that the existence of the qudit generalization of the subsystem toric code is perhaps not surprising.
However, our construction offers a radically different perspective on topological subsystem codes as constructed from (and thus closely related to) models natively defined in fewer spatial dimensions.
Such a perspective provides a new way to understand possible boundary conditions and sheds light on the single-shot QEC property that topological subsystem codes exhibit.
It also opens up possibilities of further generalizations, perhaps leading to concrete non-abelian models in three spatial dimensions.

The remainder of this article is structured as follows.
In \cref{sec:prelim}, we discuss preliminaries, including the \STC{} and quantum double models.
Then, in \cref{sec:subsystemAQD}, based on the intuition from the \STC{}, we introduce a class of subsystem abelian quantum double (\SAQD{}) codes.
In \cref{sec:errorThresholds}, we shift gears and numerically study QEC properties of \SAQD{} codes, in particular we estimate memory thresholds for phenomenological noise.
Finally, we discuss thermal stability of \SAQD{} codes in \cref{sec:gapless} and provide concluding remarks in \cref{sec:remarks}.
We also enclose a number of appendices.
In \cref{app:pauliSubsystemCodes}, we define Pauli subsystem codes, their logical operators and code parameters.
To ensure clarity, we provide a complete, and explicit, list of gauge generators for the \SAQD{} in \cref{app:explicitOperators}.
Finally, in \cref{app:gapless}, we discuss the spectral gap of the \SAQD{} Hamiltonian.


\section{Preliminaries}\label{sec:prelim}

In this section, we provide some preliminaries required for the remainder of the article.
We begin by briefly describing the \STC{}, followed by reviewing two-dimensional abelian quantum double models.

\subsection{Intuition from the subsystem toric code}\label{sec:STC}

\begin{figure}
	\centering
	\begin{subfigure}[t]{0.45\linewidth}
		\subcaption{}\label{fig:STC_A}
		\centering\includegraphicsarray{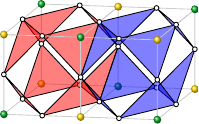}
	\end{subfigure}
	\hfill
	\begin{subfigure}[t]{0.45\linewidth}
		\subcaption{}\label{fig:STC_B}
		\centering\includegraphicsarray{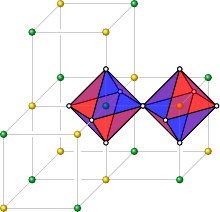}
	\end{subfigure}
	\caption{
		{\bf(a)} The three-dimensional \STC{} can be defined on the cubic lattice.
		Qubits are placed on edges.
		Pauli $X$- and $Z$-type gauge generators are depicted as triangular red and blue faces, respectively, whereas stabilizer generators are associated with volumes.
			{\bf(b)} Gauge generators can be partitioned into two commuting sets, associated to green and yellow vertices respectively.
		Around each vertex, the gauge generators form a 2-dimensional sphere supporting a toric code.
	}\label{fig:STC}
\end{figure}

Stabilizer~\cite{Gottesman1996} and subsystem~\cite{Poulin2005,Kribs2005} codes constitute two important and widely-studied classes of QEC codes.
A stabilizer code is defined by a stabilizer group $\Stab$, which is an abelian subgroup of the Pauli group $\mathcal P$ and does not contain $-I$.
By definition, logical qubits (or qudits) are encoded in the code space that is the $+1$ eigenspace of all the operators in $\Stab$.
A subsystem code is a generalization of a stabilizer code where logical information is encoded into only a subset of the logical qubits.
The remaining (i.e., unused) logical qubits are commonly called gauge qubits.
Subsystem codes can be specified by a gauge group $\mathcal G$, which can be any subgroup of $\mathcal P$.
By definition, any operator in $\mathcal G$ has a trivial action on the encoded logical information.
The stabilizer group of a subsystem code is the center of the gauge group $\mathcal G$ (up to phases).
For a detailed discussion, we refer the reader to \cref{app:pauliSubsystemCodes}.

The three-dimensional subsystem toric code is a generalization of the stabilizer toric code in an analogous way as the gauge color code is a generalization of the stabilizer color code.
A conventional way to understand both the subsystem toric code and the gauge color code is to relate them to the corresponding stabilizer codes in the same spatial dimension.
Any state in the code space of the stabilizer code can be viewed as a code state of the subsystem code
\footnote{Here, we implicitly assume that the logical subspaces of both codes are isomorphic.
	Otherwise, different logical states of the stabilizer code are mapped to subspaces corresponding to different choices of eigenvalues of stabilizers for the subsystem code.}, and a code state of the subsystem code can be transformed into a code state of the stabilizer code by fixing the gauge qubits to be in a specific state.
More formally, we have the inclusions
\begin{equation}
	\Stab \leq \Stab_{\mathrm{stab}} \leq \G,
\end{equation}
where $\Stab_{\mathrm{stab}}$ and $\S$ denote the stabilizer groups of the stabilizer and subsystem toric codes respectively, and $\G$ denotes the gauge group of the subsystem toric code.
Any state in the +1 eigenspace of $\Stab_{\mathrm{stab}}$ will be in the +1 eigenspace of $\Stab$.
Since $\Stab$ is the center of $\G$, given any operator $\sigma\in\Stab_{\mathrm{stab}}$ such that $\sigma\notin\Stab$, there must be an operator $\tau\in\G$ that anticommutes with it.
Consequently, any state in the +1 eigenspace of $\Stab$ can be transformed into a state in the +1 eigenspace of $\Stab_{\mathrm{stab}}$ by applying an operator (namely $\tau$) in $\G$, which, by definition, has no effect on the encoded information.

It is non-trivial to construct subsystem codes from the corresponding stabilizer codes, as evidenced by the amount of time between the introduction of the three-dimensional stabilizer toric code~\cite{Dennis2002} and the three-dimensional subsystem toric code~\cite{Kubica2022}.
Although the stabilizer and subsystem codes have a similar construction, they have strikingly different properties.
For example, the aforementioned subsystem codes have the single-shot QEC property, whereas the corresponding stabilizer codes do not.
And despite these subsystem code models arguably providing the most transparent illustration of the single-shot QEC property, the origins of this property, and its exact connection to self-correction, are unclear.

The \STC{} can be defined on the cubic lattice with open boundary conditions by placing qubits on the edges of the lattice, as depicted in \cref{fig:STC_A}.
By definition, each volume of the lattice supports one stabilizer generator and eight gauge generators, whose type is determined by the color associated with the volume.
The product of gauge generators around each volume is the identity operator; similarly for gauge generators surrounding each vertex.
These relations are key to the single-shot QEC property of the \STC{}, as they ensure that -1 measurement outcomes of gauge operators form closed loops.
If measurement errors cause some of the loops to be broken, then we can close them up, repairing the damage and obtaining an updated estimate for the stabilizer syndrome used in a standard QEC decoder.

Note that the gauge generators of the subsystem toric code surrounding each vertex can be viewed as the stabilizer generators of the toric code supported in a two-dimensional sphere around that vertex (\cref{fig:STC_B}).
Moreover, the whole gauge group can split into two commuting sets of operators, each of which is the collection of disjoint, two-dimensional toric codes.
This simple observation is the starting point of our construction of three-dimensional subsystem codes based on arbitrary abelian quantum double models which we present in \cref{sec:subsystemAQD}.

\subsection{Two-dimensional quantum double models}\label{sec:prelimAQD}

To any finite abelian group
\begin{align}
	G\iso \prod_{j=1}^{k}\ZZ{n_j},\label{eqn:decompAbelianGroup}
\end{align}
there is an associated topological code, called the abelian quantum double model (AQD).
The most familiar example is the toric code, which is the special case associated to $\ZZ{2}$.
For the remainder of our discussion, we specialize to the case $G\iso \ZZ{d}$ as all other cases can be obtained via stacking layers, with one layer for each factor in \cref{eqn:decompAbelianGroup}.
Let $\Lambda$ be a directed graph embedded on an orientable 2-manifold.
For simplicity, we will initially assume that $\Lambda$ is closed.
Degrees of freedom of dimension $d$ are located on the edges of $\Lambda$.
In the group basis
\begin{align}
	\mathbb{C}[\ZZ{d}]:=\Span_{\mathbb{C}}\set{\ket{g}}{g\in \ZZ{d}},
\end{align}
the generalized Pauli operators act as
\begin{subalign}
	X^{g}\ket{h} & = \ket{g+h\mod d}      \\
	Z^{g}\ket{h} & = \omega^{g h}\ket{h},
\end{subalign}
where $\omega:=\exp(2\pi i/d)$.
The commutation relations are
\begin{align}
	Z^gX^h & = \omega^{gh}X^h Z^g.
\end{align}
The Hamiltonian defining the model has the familiar form
\begin{align}
	H_{\ZZ{d}} & =-\sum_{v\in\text{vertices}} A_v-\sum_{f\in\text{faces}} B_f+h.c.\,.\label{eqn:abelianQDHamiltonian}
\end{align}
For a fixed vertex $v$, the $A_v$ term consists of $Z$ operators applied to each incoming edge of $\Lambda$, and $Z^\dagger$ to each outgoing edge.
For a fixed face $f$, the $B_f$ term consists of $X$ operators applied to each edge whose direction matches the ambient orientation, and $X^\dagger$ applied to all edges opposite to the orientation.
For example,
\begin{align}
	A_{v} & :=\includegraphicsarray{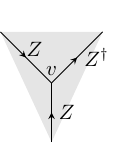},
	      &
	B_{f} :=\includegraphicsarray{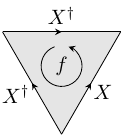},\label{eqn:abelianQDTerms}
\end{align}
with analogous action on vertices and faces of arbitrary degree.
These rules ensure that the Hamiltonian terms pairwise commute.
Ground states of the model are simultaneous $+1$ eigenstates of all Hamiltonian terms.
Interpreting a state $\ket{g}$ on an edge as the presence of a string of color $g$, the vertex terms ensure strings cannot terminate or change color (without branching), while the face terms cause a given ground state to be a superposition of all homotopic, closed loops~\cite{Levin2005}.
On a 2-sphere, the ground state is unique, while on higher genus surfaces, the ground space is degenerate.

There are two kinds of excitations above the ground state, denoted $e^g$ and $m^g$, corresponding to violating vertex and face terms respectively, which are created by local operators.
When brought close together, these particles fuse according to the group product, namely $e^g\times e^h = e^{g+h \mod d}$ and $m^g\times m^h = m^{g+h \mod d}$.
In general, any configuration of excitations on a closed manifold must fuse to vacuum $e^0m^0$ since all local operators preserve the total charge, and the ground state is neutral.
Unlike the toric code, for which $G\iso\ZZ{2}$, typical excitation configurations `branch'.
For example, whenever $G\not\iso \ZZ{2}$ we can have the following configuration of excitations
\begin{align}
	\includegraphicsarray{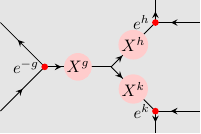}.
\end{align}

\subsubsection{Boundaries}\label{sec:AQDBoundaries}

In addition to closed manifolds, AQD models can be constructed on manifolds with boundary.
In turn, we will make use of these boundaries to construct three-dimensional subsystem codes with boundary.
For a general abelian group $G$, there are many possible gapped boundaries that can be used to modify \cref{eqn:abelianQDHamiltonian}, see for example \onlinecite{Etingof2010,Beigi2011,Kitaev2012}.
For our code construction, we will make use of the only two types of boundary that occur irrespective of the choice of $G$.
These boundaries are commonly referred to as \emph{rough} and \emph{smooth} due to their lattice implementation.

Rough boundaries are characterized by all vertices of $\Lambda$ that lie on the boundary having degree one.
The bulk face terms of the Hamiltonian are replaced by terms of the form
\begin{align}
	\tilde{B}_{f} :=
	\includegraphicsarray{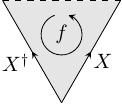},\label{eqn:QDRoughTerm}
\end{align}
where the dashed line is the physical boundary.
At smooth boundaries, all vertices have degree larger than one.
The bulk vertex terms are replaced by terms of the form
\begin{align}
	\tilde{A}_{v} :=
	\includegraphicsarray{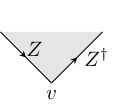},\label{eqn:QDSmoothTerm}
\end{align}

Strings of $X$ operators can terminate at rough boundaries without creating excitations, while strings of $Z$ can terminate at smooth boundaries.
Physically, this ability to remove an excitation onto the boundary is commonly called condensation.
In that language, the rough boundary condenses $e$ type excitations, while the smooth boundary condenses $m$ particles.

\subsubsection{Interfaces}\label{sec:AQDsymmetry}

Boundaries to vacuum can be understood as a particular example of an interface between two (possibly identical) AQDs.
The most familiar example is the symmetry of the toric code which exchanges the $e$ and $m$ excitations~\cite{Bombin2010}, however, even in AQD codes, they can be far richer~\cite{Bridgeman2017}.
In two-dimensional codes, the possible self-interfaces (domain walls) classify the fault-tolerant logical operators~\cite{Yoshida2015,Beverland2016,Yoshida2017,Webster2018}.

Beyond domain walls, one can also consider interfaces between domain walls~\cite{Bombin2010,Barkeshli2019,Bridgeman2020}, which can be used to further boost the computational power of two-dimensional topological codes.
We have not explored their impact on the subsystem codes we introduce here.


\section{Subsystem abelian quantum double codes in three dimensions}\label{sec:subsystemAQD}

We now proceed to define subsystem abelian quantum double (\SAQD{}) codes in three spatial dimensions.
We then explain how boundaries and interfaces can be obtained naturally from our construction.
We finish by discussing the gauge flux in the \SAQD{} codes and their relation to the qudit toric code.
For simplicity, we begin with a concrete lattice implementation, however the code can be defined more generally (\cref{sec:generalLattice}).

\begin{figure}
	\centering
	\begin{subfigure}[t]{0.45\linewidth}
		\subcaption{}\label{fig:sphereCubicLattice_a}
		\centering\includegraphicsarray{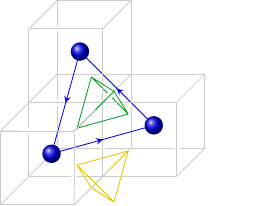}
	\end{subfigure}
	\hfill
	\begin{subfigure}[t]{0.45\linewidth}
		\subcaption{}\label{fig:sphereCubicLattice_b}
		\centering\includegraphics[width=\linewidth]{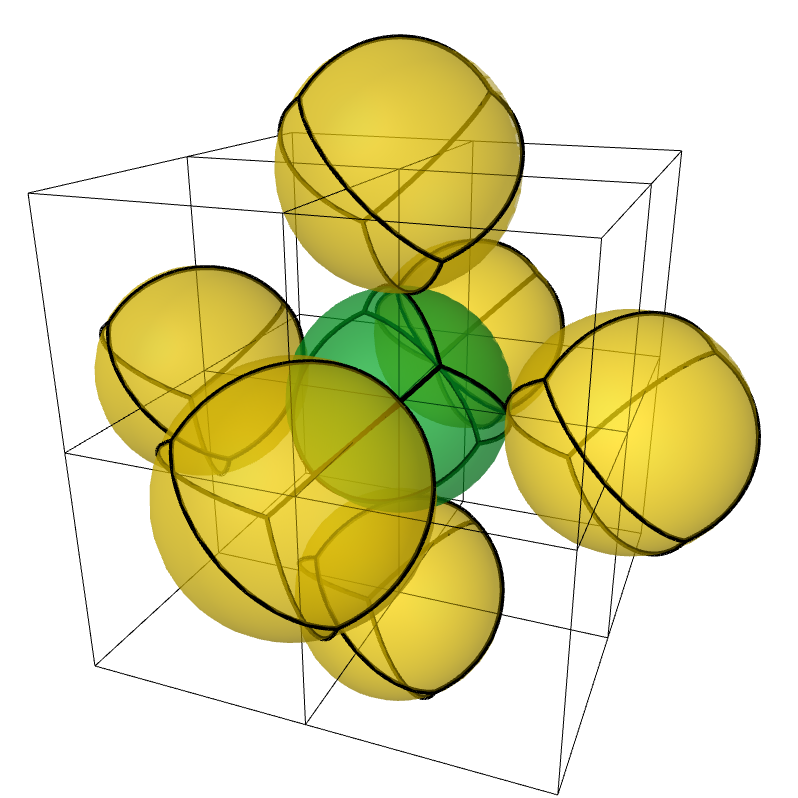}
	\end{subfigure}
	\caption{
	Given a cubic lattice with checkerboard red and blue volumes, and bicolored (green and yellow) vertices, the code lattice is defined by:
	{\bf(a)} Joining the centers of all blue volumes by directed edges.
	Code qudits are associated to the blue edges.
		{\bf(b)} Inflating the vertices to form 2-spheres until they touch, with directed triangulation induced by the blue edges.
	\interactivefig
	}\label{fig:sphereCubicLattice}
\end{figure}

\subsubsection{Bulk}\label{sec:SAQD_bulk}

Our construction of the gauge group follows the following intuition:
\begin{enumerate}
	\item Tile three-dimensional space with topological codes defined on 2-spheres.\label{enum:tile}
	\item Inflate the 2-spheres until they touch.\label{enum:inflate}
	\item Identify the degrees of freedom at the intersections.\label{enum:identify}
\end{enumerate}
\Cref{enum:tile} produces a stabilizer code which encodes no qudits.
\Cref{enum:inflate,enum:identify} convert this stabilizer code into a subsystem code.

Let $\mathcal{L}$ be the cubic lattice, with volumes colored red and blue in a checkerboard pattern, and vertices colored green and yellow in a similar pattern (see~\cref{fig:STC_A,fig:sphereCubicLattice_a}).
As mentioned in \cref{sec:prelimAQD}, defining an AQD requires a directed graph.
The volume coloring is an auxiliary tool enabling consistent definition of these graphs throughout the code as follows.
Connect the centers of nearest neighbor blue volumes of $\mathcal{L}$ with directed edges, as depicted in \cref{fig:sphereCubicLattice_a}, with the direction of each edge chosen arbitrarily.
Qudits of the final code ultimately reside on the added, blue edges~\footnote{To visualize the lattice, it may be helpful to note that the blue edges are in 1:1 correspondence with the edges of the original cubic lattice.}.
With this auxiliary lattice in place, we are in a position to define the \SAQD{} gauge group.

At each vertex of $\mathcal{L}$, we place a microscopic (i.e., not scaling with code size) sphere, which we imagine to be hosting a copy of the appropriate AQD (\Cref{enum:tile}).
As in \cref{fig:sphereCubicLattice_a}, the auxiliary blue lattice induces a directed triangulation on these spheres, allowing an explicit lattice model.
In addition to the directed graph, it is necessary to provide an orientation to the 2-spheres, which can be achieved by defining a vector from green vertices of $\mathcal{L}$ to yellow vertices.
We now inflate these spheres until their triangulations intersect with the auxiliary lattice (\Cref{enum:inflate}).
Each sphere is split by six directed edges into four triangular regions and thus is homeomorphic to the boundary of a tetrahedron.
Finally, the qudits on the common edges are identified (\Cref{enum:identify}), leading to the geometry depicted in \cref{fig:sphereCubicLattice_b}.
Qudits reside on the dark black edges which form the triangulation in \cref{fig:sphereCubicLattice_b}.

With the geometry established, we aim to use the topological features of the AQD code to ensure the resulting code has the single-shot QEC property.
The gauge group of the resulting code is given by
\begin{align}
	\G_{\ZZ{d}}:= & \grouppresentation{A_v,B_f}{v,f\in\text{all 2-spheres}},\label{eqn:bulkGaugeGroup}
\end{align}
where $A_v,\, B_f$ are defined in \cref{eqn:abelianQDTerms}.
Gauge operators of different colors do not commute, so this fails to be a stabilizer code.
Although the model can no longer be thought of as a collection of unbroken loops as discussed in \cref{sec:prelimAQD}, it remains true that the total charge on a sphere must be trivial.

In the remainder of this section, we discuss the stabilizers of the code, and how boundaries can be introduced.
As in the \GCC{} and \STC, the latter are necessary to ensure a non-trivial codespace.
We show numerical evidence that the code has the single-shot property in \cref{sec:errorThresholds}.

A full accounting of the gauge generators is provided in \cref{app:explicitOperators} for completeness.

\subsubsubsection{Stabilizers}

\begin{figure}
	\centering
	\begin{subfigure}[t]{0.45\linewidth}
		\subcaption{}\label{XStabilizer_G}
		\centering\includegraphics[width=.7\linewidth]{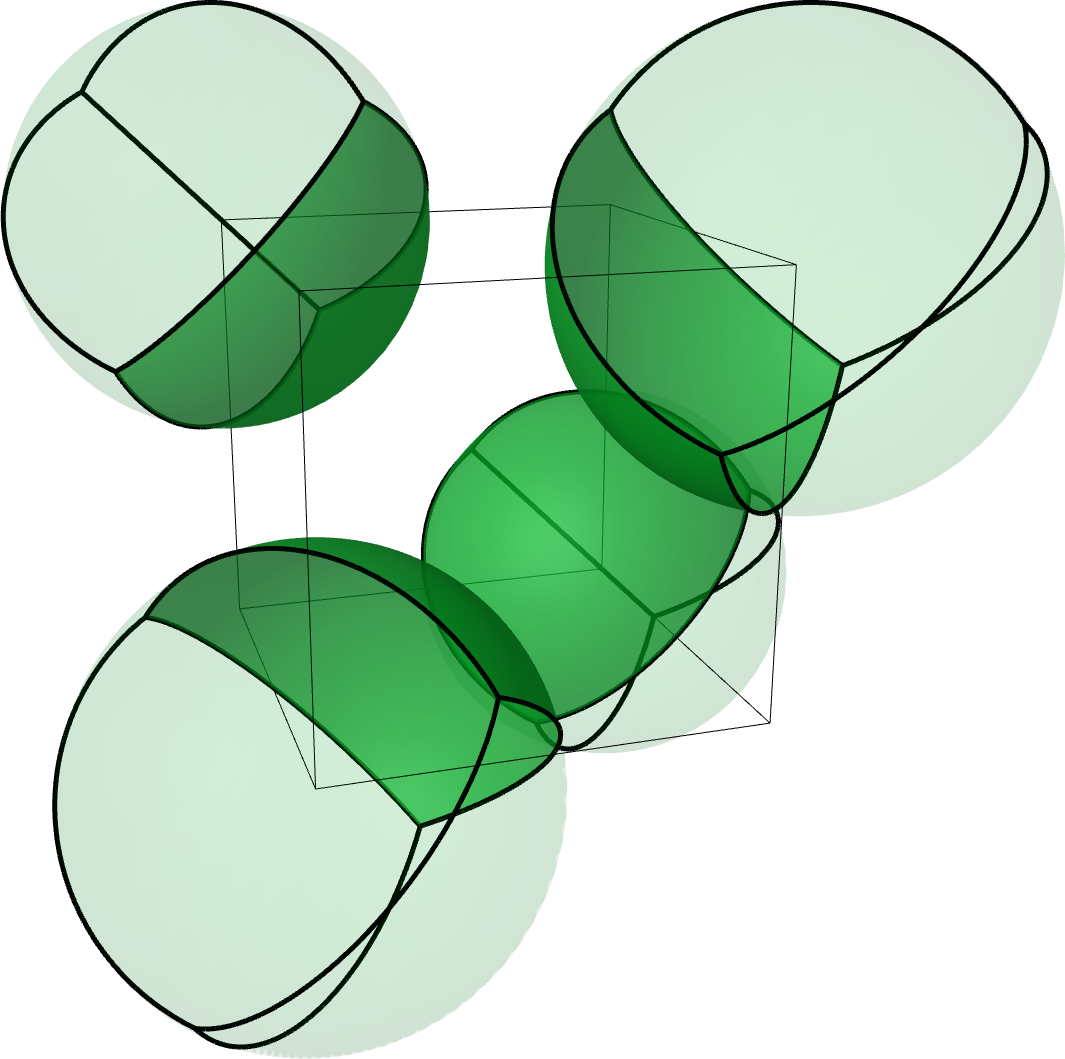}
	\end{subfigure}
	\begin{subfigure}[t]{0.45\linewidth}
		\subcaption{}\label{XStabilizer_Y}
		\centering\includegraphics[width=.7\linewidth]{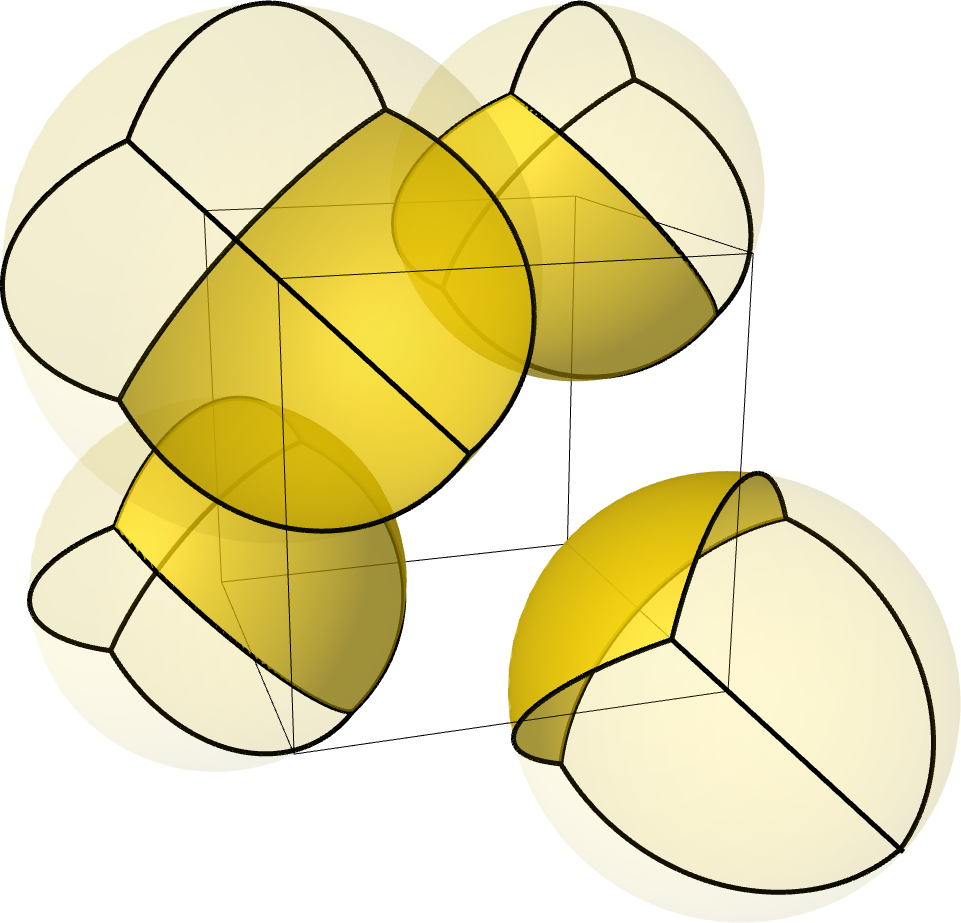}
	\end{subfigure}
	\begin{subfigure}[t]{0.95\linewidth}
		\subcaption{}\label{fig:XStabilizer}
		\centering\raisebox{2mm}[18mm]{\makebox[\textwidth][c]{\includegraphicsarray{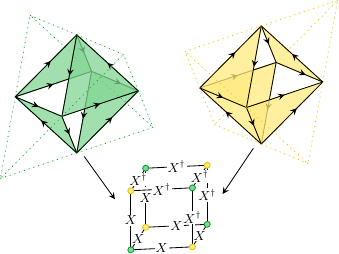}}}
	\end{subfigure}
	\begin{subfigure}[t]{0.95\linewidth}
		\subcaption{}\label{fig:ZStabilizer}
		\centering\includegraphicsarray{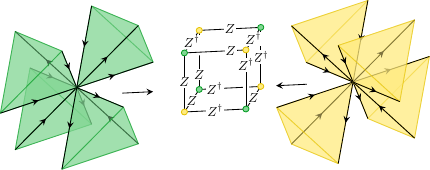}
	\end{subfigure}
	\caption{%
		Stabilizer operators are those that can be recovered from gauge generators of either color.
		They are associated to the volumes of the cubic lattice.
			{\bf(a)} and {\bf(b)} show the faces participating in a given $X$-type stabilizer.
		In {\bf(c)} and {\bf(d)}, we show how to recover the stabilizers by multiplying certain gauge generators in two different ways.
		\interactivefig
	}\label{fig:bulkStabilizers}
\end{figure}

Since the underlying topological codes are commuting, and distinct spheres of the same color are non-intersecting, the stabilizer group corresponding to the gauge group \cref{eqn:bulkGaugeGroup} is given by any operator that can be created from generators of either color.
One can verify that locally generated stabilizers are associated with the volumes of the cubic lattice.
Recall that we began with as bicolored cubic lattice $\mathcal{L}$.
Red volumes only contain faces of tetrahedra, such as that in \cref{fig:XStabilizer}, and consequently support $X$-type stabilizer operator.
Conversely, blue volumes contain vertices of tetrahedra such as those depicted in \cref{fig:ZStabilizer}, and support $Z$-type stabilizers.

In addition to the local stabilizers, the model on the 3-torus ($\Torus{3}$) supports sheet-like stabilizers in each coordinate plane as depicted in \cref{eqn:sheetStabilizers}.
By introducing boundaries, these stabilizer operators can be promoted to logical operators.

\subsubsection{Boundaries}\label{sec:boundaries}

\begin{figure}\centering
	\begin{subfigure}[t]{0.4\linewidth}
		\subcaption{}\label{fig:T2xI}\vspace*{-5mm}
		\centering\includegraphicsarray{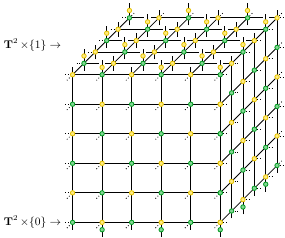}
	\end{subfigure}
	\hfill
	\begin{subfigure}[t]{0.4\linewidth}
		\subcaption{}\label{fig:T2xI_top}
		\centering\includegraphicsarray{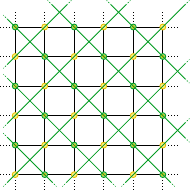}
	\end{subfigure}
	\\
	\begin{subfigure}[t]{0.49\linewidth}
		\subcaption{}\label{fig:T2xI_ops}
		\centering\includegraphicsarray{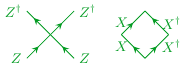}
		\\
		\subcaption{}\label{fig:axes}\vspace*{-5mm}
		\centering\includegraphicsarray{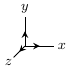}
	\end{subfigure}
	\hfill
	\begin{subfigure}[t]{0.49\linewidth}
		\subcaption{}\label{fig:T2xI_log}\vspace*{-5mm}
		\centering\includegraphicsarray{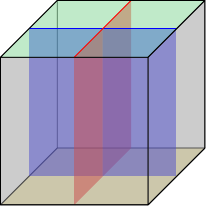}
	\end{subfigure}
	\caption{
		$\Torus{2}\times\Interval$ with class \ref{enum:macroBnd} boundaries.
		2-dimensional AQD models are on the top and bottom boundaries, with lattice periodic in other directions.
			{\bf(b)} shows the view from above.
		All spheres incident on the top boundary are colored yellow, and the AQD model on the surface is colored green.
		Colors are reversed at the bottom boundary.
			{\bf(c)} Gauge generators assigned to the top surface, realizing a class \ref{enum:macroBnd} boundary.
		To aid discussion, we fix the coordinate axes in {\bf(d)} for the remainder of the manuscript.
			{\bf(e)} Logical operators on the \SAQD{} are sheet-like operators that intersect the boundary.
		They can be dressed to string-like operators supported on the boundary, corresponding to the logical operators of the boundary AQD.}\label{fig:openTopLattice}
\end{figure}

Due to the sheet-like stabilizer operators on a closed manifold, boundaries are required to have a non-trivial codespace
This is also seen in the subsystem toric and gauge color codes.
This suggests that their nature, although not truly topological, is closely related to models in one fewer spatial dimension.

We will discuss two classes of boundaries for the \SAQD{}, both inherited from the underlying AQD:
\begin{enumerate}\renewcommand{\theenumi}{\Roman{enumi}}%
	\item Boundaries given by macroscopic copies of the underlying AQD (as opposed to the microscopic 2-spheres at the vertices of $\mathcal{L}$). \label{enum:macroBnd}
	\item Boundaries of the subsystem code induced by the one-dimensional boundaries of the AQD.\label{enum:1DBnd}
\end{enumerate}

The type of boundary (Class \ref{enum:macroBnd}) we consider is obtained by defining the code on $\Torus{2}\times \Interval$, where $\Interval:=\I$, as shown in \cref{fig:T2xI}.
Gauge generators associated to the boundaries $\Torus{2}\times\{0\}$ and $\Torus{2}\times\{1\}$ (referred to as bottom and top respectively) are those of the underlying AQD model.
Unlike the spheres in the bulk, these surfaces are macroscopic (\cref{fig:T2xI_top}).
To encode qudits, it is necessary to color the boundaries top and bottom oppositely.
As in \onlinecite{Kubica2022}, we need to introduce additional qudits as discussed in \cref{app:explicitOperators} to make this possible.

Orientation of the top and bottom surfaces is defined in the same way as the bulk spheres (\cref{sec:SAQD_bulk}).
The orientation vectors point from green to yellow.
As an example, the top surface, colored green, has the gauge generators shown in \cref{fig:T2xI_ops}.

\subsubsubsection{Open boundaries}

\begin{figure}[t]
	\begin{subfigure}[t]{0.48\linewidth}
		\subcaption{}\label{fig:sphereIntersectionRough}\vspace*{-5mm}
		\centering\includegraphics[height=45mm]{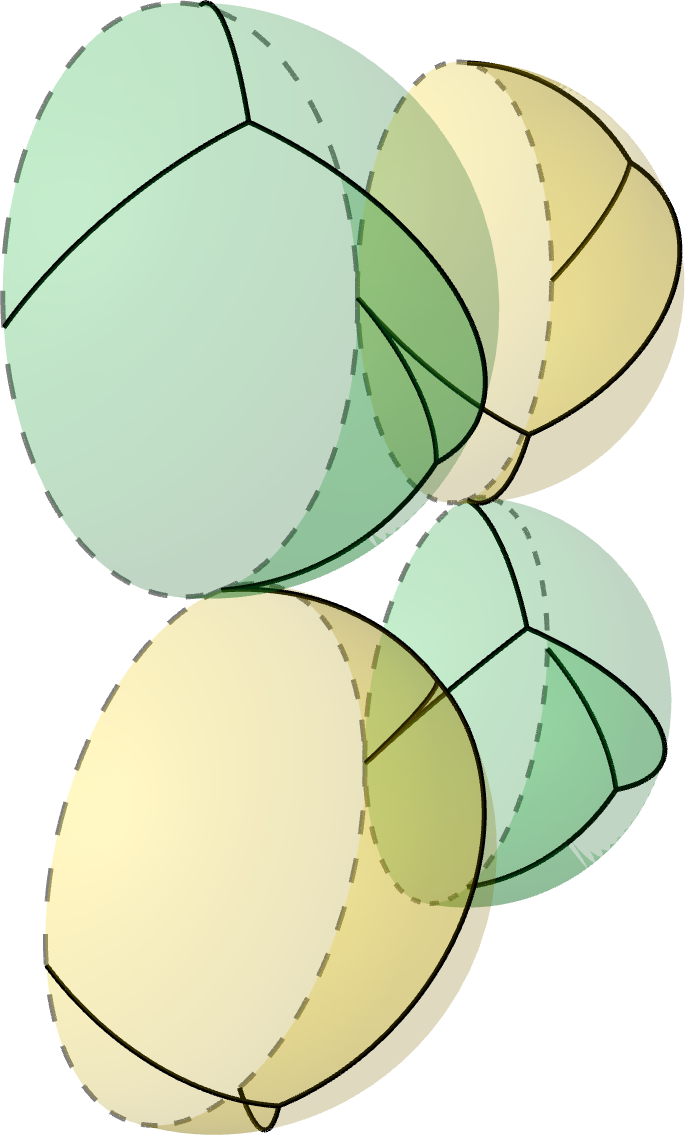}
	\end{subfigure}
	\hfill
	\begin{subfigure}[t]{0.48\linewidth}
		\subcaption{}\label{fig:sphereIntersectionSmooth}\vspace*{-5mm}
		\centering\includegraphics[height=45mm]{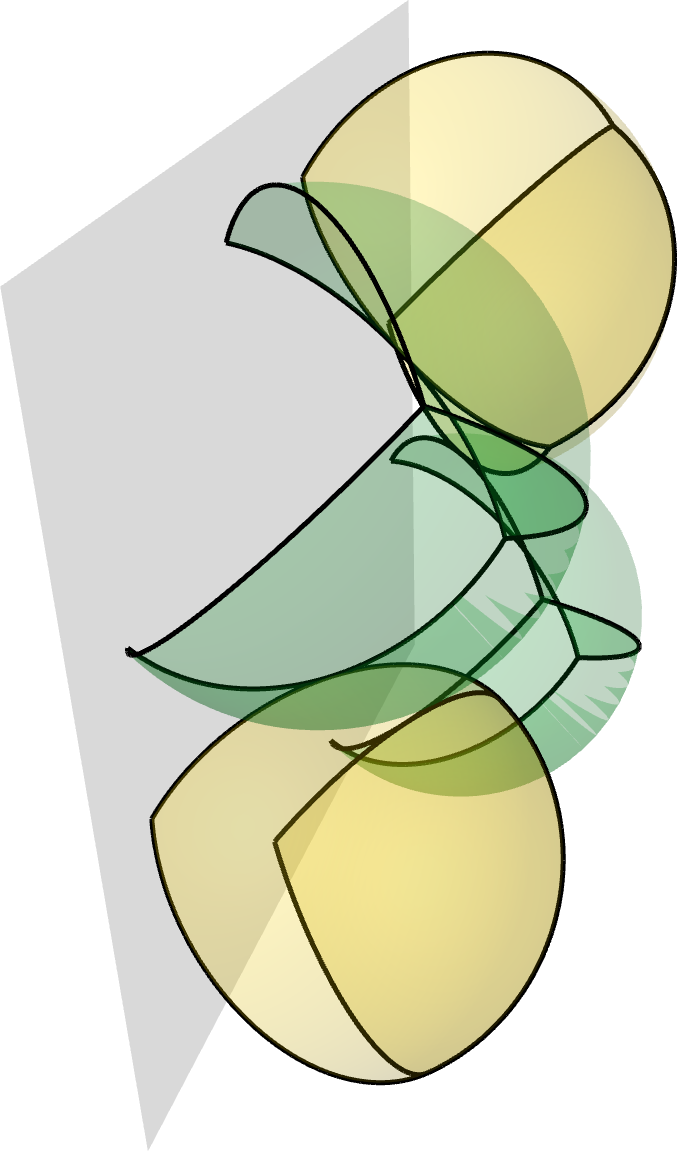}
	\end{subfigure}
	\caption{Class \ref{enum:1DBnd} boundaries.
		At the rough {\bf(a)} and smooth {\bf(b)} boundaries, the half spheres are decorated with a quantum double model with the appropriate type of boundary.
		\interactivefig
	}\label{fig:RSBoundarySpheres}
\end{figure}

The other type (Class \ref{enum:1DBnd}) of boundaries we consider is obtained by introducing boundaries on the 2-spheres of AQD, as introduced in \cref{sec:AQDBoundaries}.
This allows definition of the code on more general manifolds (with boundaries), such as the cube.
For the remainder of this section, the top and bottom boundaries will be macroscopic AQD models.

As discussed in \cref{sec:AQDBoundaries}, all AQD models can be equipped with rough and smooth boundaries.
These boundaries of the two-dimensional models can be used to define boundaries of the associated three-dimensional \SAQD{} code.
When a green or yellow sphere intersects with a boundary, we can choose to equip the code on that sphere with rough boundaries, as shown in \cref{fig:sphereIntersectionRough}.
We refer to such a boundary as a rough boundary of the \SAQD{}.
Conversely, we could choose smooth boundaries for all spheres that intersect the boundary, as in \cref{fig:sphereIntersectionSmooth}, defining a smooth boundary for the \SAQD{}~\footnote{%
	For clarity, we emphasize that these boundaries are not the same as those that would be obtained by defining the three-dimensional stabilizer toric code on the same lattice; rather, they are named in terms of the two-dimensional model.
}.

At the $y$-edges of the cube, where different boundaries meet, the spheres carry mixed boundary conditions.
Where the rough/smooth boundaries intersect the top and bottom boundaries, the macroscopic AQD inherits the appropriate class of boundary.

\subsubsubsection{Three-body model}

By construction, all bulk gauge generators, and those associated to rough and smooth boundaries have non-trivial support on at most three qudits.
The macroscopic AQD models on the top and bottom boundary introduces four body terms, however this is not essential.
It may be advantageous from an error correction perspective to reduce these to weight three~\cite{Dennis2002}.
In \cref{sec:threeBodyOnly}, following \onlinecite{Kubica2022}, we show how these boundaries can be modified to reduce the weight.

\subsubsection{Logical operators}\label{sec:logical}

When defined on $\Torus{3}$, the \SAQD{} code has both local and macroscopic (sheet-like) stabilizer operators.
Recall that the purpose of introducing boundaries was to promote these sheet-like stabilizers to logical operators.
Sheets that do not intersect the boundary remain as stabilizers.
For example, on $\Torus{2}\times\Interval$, the sheet-like operators in the $xz$-plane (\cref{fig:axes}) are in $\G_{\ZZ{d}}$, so remain stabilizers.
Conversely, sheet-like operators that do intersect the boundary can no longer be constructed from gauge generators, but remain in the centralizer of the gauge group.
On $\Torus{2}\times\Interval$, this gives sheet-like logical operators in both the $xy$ and $yz$-planes, meaning the code supports two logical qudits.

In the bulk, these bare logical operators can be dressed with either green or yellow gauge generators to reduce their weight.
Although these dressed logical operators no longer lie in the centralizer of the gauge group, they do commute with the stabilizer, and so remain good logical operators.
By dressing the sheet-like logical operators with green generators, the sheet is reduced to a string-like operator on the green boundary, and vice versa.
These strings correspond to the logical operators of the boundary AQD as shown in \cref{fig:T2xI_log}.
For a more explicit discussion of the logical operators we refer the reader to\cref{app:explicitOperators}.

\subsubsection{Code parameters}

On a closed manifold, \SAQD{}s do not encode any qudits due to the macroscopic stabilizer generators.
When placed on a manifold $M\times \Interval$ (possibly where $M$ has boundaries), with AQD models at the top and bottom, the codes have a number of logical qudits equal to the number for that AQD model on $M$.

For the manifolds discussed above, the code parameters are summarized in \cref{tab:codeParameters}, and obtained in \cref{app:explicitOperators}.

\begin{table}
	\centering
	\renewcommand{\arraystretch}{1.25}
	\begin{tabular}{
		!{\vrule width 1pt}>{\columncolor[gray]{.9}[\tabcolsep]}c!{\vrule width 1pt}c!{\vrule width 1pt}c!{\vrule width 1pt}c!{\vrule width 1pt}c!{\vrule width 1pt}}
		\specialrule{1pt}{0pt}{0pt}
		\rowcolor[gray]{.9}[\tabcolsep] Manif. & $ n $                 & $ k $ & $ \bareDistance $ & $ \dressedDistance $ \\
		\specialrule{1pt}{0pt}{0pt}
		$ \Torus{3} $                          & $ 3L^3 $              & $ 0 $ & $ - $             & $ - $                \\
		$ \Torus{2}\times \Interval $          & $ 3 L^3+2 L^2 $       & $ 2 $ & $ 2L(L+1) $       & $ L $                \\
		$ (\Torus{2}\times \Interval)^\prime $ & $ 3 L^3+2 L^2 + L^2 $ & $ 2 $ & $ 2L(L+1) $       & $ L/2 $              \\
		$ \Interval^3 $                        & $ 3 L^3+6 L^2+5 L+1 $ & $ 1 $ & $ 2 L^2+3 L+1 $   & $ L+1 $              \\
		\specialrule{1pt}{0pt}{0pt}
	\end{tabular}
	\caption{%
		Code parameters for the \SAQD{}s on various manifolds of linear size $L$, possibly with boundary.
		The cube ($\Interval^3$) is equipped with rough/smooth boundaries, while $ (\Torus{2}\times \Interval)^\prime $ is the code with only 3-body generators as discussed in \cref{sec:boundaries}.
		The number of physical qudits is denoted $n$, while the number of logical qudits is $k$.
		The bare (dressed) distance is denoted $\bareDistance$ ($\dressedDistance$).
	}\label{tab:codeParameters}
\end{table}

\subsubsection{Interfaces}\label{sec:interfaces}

As discussed in \cref{sec:AQDsymmetry}, locality-preserving gates on topological codes in two spatial dimensions are in one-to-one correspondence with codimension-1 interfaces.
Such gates can be lifted to the associated \SAQD{} code, simply by applying the appropriate circuit to each microscopic or macroscopic copy of the AQD.
Such an action will locally preserve the full gauge group, meaning its action can be restricted straightforwardly to the boundaries.
Since the action on the dressed logical operators is identical to that in the input code, the logical action is also identical.
Conversely, any non-Clifford gate that can be enacted on these \SAQD{} codes cannot be a symmetry of the full gauge group in this way~\cite{Bravyi2013b,Pastawski2015}.%

In this work, we have not investigated the interpretation of higher codimension defects, such as point-like defects in two dimensions.
This class includes both anyonic excitations, and more exotic objects such as twist defects~\cite{Bombin2010}.

\subsubsection{General lattice implementation}\label{sec:generalLattice}

In this section, we have restricted our discussion to the code defined on the cubic lattice, possibly with boundaries.
As in \onlinecite{Kubica2022}, the definition can be extended to any \define{colorable octahedral lattice} (see \onlinecite{Kubica2022} for the definition).
Given such a lattice, we can connect the blue vertices as in \cref{fig:sphereCubicLattice} and inflate the green/yellow vertices to construct the code.

\subsection{Physics of the gauge flux in the \SAQD{} codes}\label{sec:errorConfigs}

\begin{figure}
	\begin{subfigure}[t]{0.45\linewidth}
		\subcaption{}\label{fig:gaugeFluxClean}\vspace*{-5mm}
		\centering\includegraphicsarray{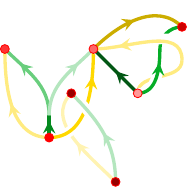}
	\end{subfigure}
	\hfill
	\begin{subfigure}[t]{0.45\linewidth}
		\subcaption{}\label{fig:gaugeFluxDirty}\vspace*{-5mm}
		\centering\includegraphicsarray{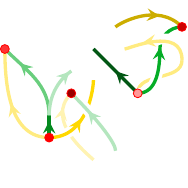}
	\end{subfigure}
	\caption{
		Gauge flux configurations.
		Unlike the \STC{}, but similarly to the \GCC{}, gauge flux can split for general \SAQD{}s.
		In addition, the flux lines can carry $g\in \G$ units of flux (indicated by shading).
			{\bf(a)} Typical gauge flux configuration in the absence of measurement errors.
		Flux obeys a Gauss law defined by the input AQD model.
			{\bf(b)} When measurement errors occur, flux lines do not obey the required Gauss law.
		\interactivefig}
	\label{fig:gaugeFlux}
\end{figure}

We now briefly discuss the behavior of the gauge flux in the three-dimensional \SAQD{} codes (which is qualitatively similar to that of the gauge color code~\cite{Bombin2018,Beverland2021}).
For simplicity, we restrict our attention to the $X$-type gauge operators $B_f$, in turn, this means we will only discuss red ($X$-type) stabilizer operators.
We will refer to an outcome $\omega^g$ when measuring $B_f$, as a \define{gauge flux} of $g$ flowing into the associated sphere.
A flux of $g^{-1}$ flowing in can be interpreted physically as $g$ flowing out.
There are two colors of flux lines, green flows into green vertices and yellow flux flows into yellow vertices.
Just like in the gauge color code, flux lines can split and fuse according to the product on the group $G$.

\subsubsubsection{No errors}
Since the gauge group is non-abelian, the outcome of any $B_f$ measurement appears random.
This means that on any measurement round, the gauge flux configuration will be random, however there are relations:
\begin{enumerate}
	\item Due to the topological order on the 2-spheres at each vertex, there can be no net charge on the sphere, and so the total flux in to any vertex must be $1_G$.\label{enum:nocharge}
	\item Since every stabilizer operator can be formed from either only green $B_f$ or only yellow $B_f$, the total green flux leaving any red volume must equal the yellow flux leaving the same volume.\label{enum:flux}
	\item By definition, the code space has all stabilizer operators equal to $+1$, so the total flux of either color leaving each red volume must be $1_G$.
\end{enumerate}
The first two relations, referred to as Gauss laws, are required to hold by construction, with the first being due to the two-dimensional topology and the second due to the three-dimensional geometry.

\subsubsubsection{Physical errors}
When there are physical errors acting on the code, the first two relations must still hold.
Errors can be thought of as creating $G$ charge at the red volumes, so can source flux, leading to violation of the third relation (\cref{fig:gaugeFluxClean}).
This is then used to detect, and with the aid of a decoder to correct errors.

\subsubsubsection{Measurement errors}

When measurement errors occur, the resulting gauge flux configurations may violate any of the relations, as in \cref{fig:gaugeFluxDirty}.
Since relations [\ref{enum:nocharge}] and [\ref{enum:flux}] are necessarily true in the physical system, observed charge at the vertices, or mismatch of the yellow and green charge at a volume can be used to correct the defective measurements.

We discuss how the gauge flux can be used to correct both types of error in \cref{sec:errorThresholds}.

\subsection{Relation to qudit toric codes}\label{sec:quditToricCodes}

We conclude this section by briefly discussing the relationship between the \SAQD{} codes and the qudit toric code in three spatial dimensions.
For simplicity, we restrict our attention to the 3-torus.

Recall from \cref{fig:bulkStabilizers} that the stabilizer group associated to the \SAQD{} code is generated by operators associated to the volumes of the cubic lattice.

We can define another abelian subgroup of $G$ via \emph{gauge fixing}~\cite{Paetznick2013,Anderson2014,Bombin2015}.
This technique allows new stabilizer codes to be constructed from subsystem codes by choosing a maximal subset of commuting, local gauge generators to be the stabilizer group.
Consider the subgroup of $\G_{\ZZ{d}}$ as the group generated by all locally generated stabilizers (associated to volumes) of $Z$-type and all $X$-type gauge operators.
On the blue lattice (\cref{fig:sphereCubicLattice_a}), this gives $Z$-type stabilizers associated to vertices, and $X$-type stabilizers associated to faces.
The code defined in this way deserves to be called the $\ZZ{d}$ toric code, since it reduces to the standard construction when $d=2$.
To the best of our knowledge, such a definition has not appeared explicitly in the literature.


\section{Error thresholds and decoding}\label{sec:errorThresholds}

In this section, we use Monte Carlo simulations to estimate the error threshold of the \SAQD{} codes defined for $\mathbb{Z}_d$ and supported on $\Interval^3$ (with rough/smooth boundaries as discussed above) as a function of $d$.
The simulation code is available on GitHub~\cite{GithubRepo}.

\subsection{Setup and error model}\label{sec:errorModel}

The \SAQD{} codes for $\mathbb{Z}_d$ are CSS~\cite{Calderbank1996,Steane1996a}, meaning that we can choose a generating set for the gauge group where each generator consists of either Pauli $X$-type operators or Pauli $Z$-type operators only.
Consequently, we can decode Pauli $X$- and $Z$-type errors independently and we therefore restrict our attention to $Z$-type errors.
Specifically, we consider single-qudit error channels of the following form
\begin{equation}
	\mathcal E (\rho) = (1-p) \rho + \frac{p}{d-1} \sum_{j=1}^{d-1} Z^j \rho Z^{-j}.
	\label{eq:quditNoise}
\end{equation}
Error channels of this form have been used previously in studies of qudit topological codes~\cite{DuclosCianci2013,DuclosCianci2014,Anwar2014,Watson2015,Hutter2015}.

We define a QEC cycle to consist of qudit $Z$-type errors, noisy $X$-type gauge flux measurements and the application of a recovery operator produced by a decoding algorithm (that only uses measurement outcomes from that QEC cycle).
We simulate the above with the following sequence of steps.
\begin{enumerate}
	\item Update the residual error by applying a $Z$-type gauge operator chosen uniformly at random from the full gauge group (not just from a set of generators).\label{enum:randomZGaugeOp}
	\item Update the residual error by applying the error channel in~\cref{eq:quditNoise} to each qudit in the model.
	\item Calculate the $X$-type gauge flux.
	\item \label{enum:noisyflux} For each flux measurement, with probability $p$ return the outcome chosen uniformly at random from $\mathbb{Z}_d \setminus \{ x\}$, where $x \in\mathbb{Z}_d$ is the correct outcome.
	\item Update the residual error with the qudit recovery operator output by the decoder described in \cref{sec:clusteringDecoder}.
\end{enumerate}
We apply the random gauge operator in \cref{enum:randomZGaugeOp} above as we are implicitly assuming alternating rounds of $X$-type and $Z$-type gauge operator measurements.
We begin the simulation by initializing the residual error to be trivial and we then simulate $t$ QEC cycles.
After the final QEC cycle we model readout by a round of ideal (no measurement error) decoding.

\subsection{Clustering decoder}
\label{sec:clusteringDecoder}

We break the decoding problem into two stages: (a) flux validation and (b) qudit correction.
In the first stage we attempt to correct the measurement error using the local relations between gauge operators.
And in the second stage we attempt to find a (qudit) recovery operator using the stabilizer eigenvalues, which we construct from the corrected flux measurements produced in the first stage.
Both decoding stages can be captured by the following decoding problem.
Let $H \in \mathbb F_d^{m \times n}$ be a parity-check matrix, where each row is a parity check and the each column corresponds to a variable.
We are given an error syndrome $\sigma \in \mathbb F_d^m$ and our task is to output a recovery operator $y \in \mathbb F_d^n$ such that $H y = \sigma$ (and we may further require that $y$ has minimal Hamming weight).
In the first decoding stage the variables are the flux measurement outcomes and the checks are the local relations, while in the second stage the variables are the qudits and the checks are the stabilizers.

In the $\ZZ{2}$ case, both decoding problems can be solved efficiently and exactly by the minimum-weight perfect matching algorithm~\cite{Kubica2022}, as in this case each variable is in the support of at most two parity checks.
However, for qudits of local dimension $d > 2$ the generalization of matching is inefficient and we therefore use a clustering decoder~\cite{bravyi2013a,DuclosCianci2013}.
We note that two-stage decoders such as the one we have described have previously been used to decode a variety of quantum codes with the single-shot property~\cite{Brown2016a,Duivenvoorden2019,Kubica2022}.

To explain how the clustering decoder works, we review the construction of a syndrome graph $\graph$ from a parity-check matrix $H$ with the property that each column contains at most two nonzero entries (the parity-check matrices for both decoding stages have this property).
For each row of $H$ we create a node in $\graph$ and we connect two vertices if their corresponding rows in $H$ have overlapping support.
We also create an auxiliary node (representing the physical boundary of the model) and for any column with only one nonzero entry we connect the corresponding node the auxiliary node.

A given error syndrome $\sigma$ can be interpreted as specifying the charge present at each node in $\graph$.
We define a cluster to be a subgraph of $\graph$ with the structure of a tree.
The charge of a cluster is the product (in $G$) of the charges of its nodes.
We define a cluster to be neutral if and only if either its charge is $1_G$ or it contains the auxiliary node (representing the physical boundary).
The first step of the clustering decoder is to initialize a list of clusters, where there is a cluster for each of the nodes in $\graph$ with non-trivial charge.
The decoder then iterates the following three subroutines until the list of clusters is empty.
\begin{enumerate}
	\item \textbf{Grow}:
	      For each non-neutral cluster, iterate over its leaves as follows.
	      For each neighbor of the leaf, check if the neighbor is in the cluster.
	      If not, add it to the cluster and:
	      {\bf(i)} If the neighbor is the auxiliary node then stop growing the cluster and mark it as neutral.
		      {\bf(ii)} If the neighbor belongs to a different non-neutral cluster, then mark the clusters to be merged.
	      If the combined charge of the clusters to be merged is neutral then stop growing the current cluster.
	\item \textbf{Merge}:
	      For each pair of clusters marked for merging, attach the smaller cluster as a subtree of the larger cluster.
	\item \textbf{Neutralize}:
	      For each neutral cluster compute a correction by moving all the nonzero charges to the root (if the charge of the cluster is trivial) or otherwise by moving all the nonzero charges to the auxiliary node.
	      After neutralizing a cluster, remove it from the list of clusters.
\end{enumerate}

We remark that the performance of our clustering decoder can be further improved by optimizing the ``Growth'' and ``Merge'' steps.
In particular, we could adapt them to resemble the corresponding steps of the union-find decoder~\cite{Delfosse2021}.
Given that the path from each charge to the root of its cluster can be updated during the ``Growth'' and ``Merge'' steps, the runtime analysis for the union-find decoder is still applicable, therefore implying that our optimized decoder can be implemented in almost linear time.
We emphasize however that the standard version of the union-find decoder is only applicable in our setting for the $\ZZ{2}$ model.

\subsection{Simulation results}\label{sec:simResults}

\begin{figure*}
	\centering\includegraphics[width=.9\linewidth]{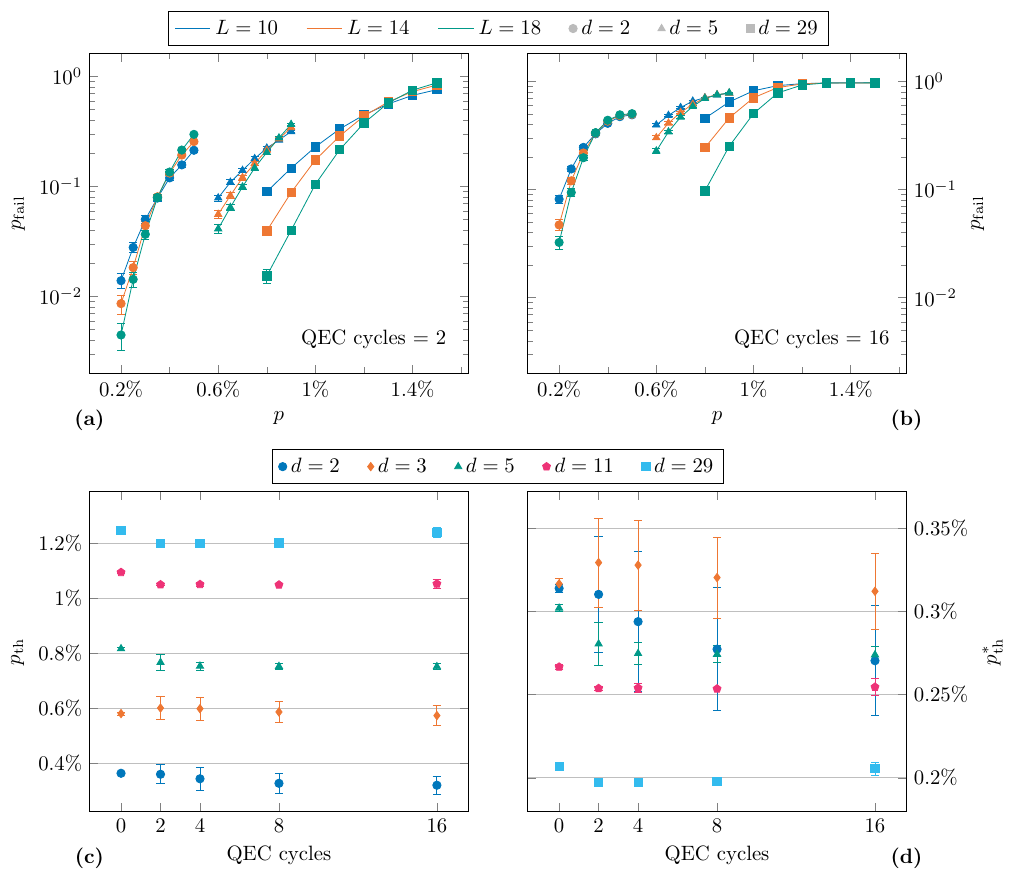}
	{\phantomsubcaption\label{fig:logicalFailRate_2}}
	{\phantomsubcaption\label{fig:logicalFailRate_16}}
	{\phantomsubcaption\label{fig:threshold}}
	{\phantomsubcaption\label{fig:threshold_rescale}}
	\caption{
		{\bf{(a)}}
		A plot of the logical error rate, $p_{\mathrm{fail}}$, as a function of the qudit and measurement error rate, $p$, over $t= 2$ QEC cycles for various $\ZZ{d}$ \SAQD{}s.
		The error bars show the Agresti-Coull 95\% confidence intervals~\cite{DasGupta2001}.
		For each value of $d$, the error threshold is the point where the curves for different $L$ intersect.
			{\bf{(b)}}
		A plot analogous to {\bf{(a)}} but for $t=16$ QEC cycles.
		Although the logical error rates increase, the error thresholds are essentially the same.
			{\bf{(c)}}
		A plot of the error thresholds, $p_{\mathrm{th}}$, as a function of the number of QEC cycles for different values of $d$.
		The error thresholds are estimated from data such as those shown in {\bf{(a)}} and {\bf{(b)}} using finite-size scaling analysis~\cite{Harrington2004}, and the error bars are derived using bootstrapping with 100 estimates.
			{\bf{(d)}}
		A plot of the rescaled error thresholds, $p_{\mathrm{th}}^*$, as a function of the number of QEC cycles, where the error thresholds from {\bf{(c)}} are rescaled according to \cref{eq:quditPScaling}.
		In this case the advantage for larger $d$ disappears.
	}
	\label{fig:numerics}
\end{figure*}

We investigate the behavior of the error threshold of the model as a function of the local dimension $d$ and the number of QEC cycles $t$.
We find that the error threshold is insensitive to the number of QEC cycles, but increases with the local dimension $d$, in agreement with previous results~\cite{DuclosCianci2013,DuclosCianci2014,Anwar2014,Watson2015,Hutter2015}.
The results are shown in \cref{fig:numerics}.

We remark that the error channel in \cref{eq:quditNoise} may bias the results in favor of large $d$, as we are squeezing an ever larger number of error events into the same probability interval.
As an alternative, we could consider a model, where $p$ increases with $d$.
For example, we can view $m$ qubits as an effective qudit of dimension $d= 2^m$.
According to \cref{eq:quditNoise} the probability of no error is $1-p(d)$, where we have now included a dependence on $d$.
On the other hand, the probability of no error for $m$ qubits is $(1 - p(2))^m$.
Equating these two values gives
\begin{equation}
	p(d) = 1 - (1 - p(2))^{\log_2 d} \approx p(2) \log_2 d.
	\label{eq:quditPScaling}
\end{equation}
In \cref{fig:threshold_rescale}, we rescale the error thresholds according to the above equation, which dramatically changes the results.
There is a clear degradation of performance for large $d$, giving the opposite conclusion to that of \cref{fig:threshold}.
This suggests that the increased performance of qudit topological codes for large $d$ reported in previous works may in fact be an artifact of the error model used in simulations.

The error threshold we find for the $\ZZ{2}$ case is very close to the error threshold found for the gauge color code in Ref.~\cite{Brown2016a} using a similar decoder.
However in that case the error threshold decays from a higher starting value whereas we observe no meaningful change in the error threshold as we vary the number of QEC cycles (we discuss a possible reason for this effect in \cref{sec:decoderComparison}).

\subsection{Decoder comparison}
\label{sec:decoderComparison}

For the $\ZZ{2}$ model, we compare the performance of four different decoders, where for each of the two decoding stages we use either the clustering decoder described in Ref.~\cref{sec:clusteringDecoder} or a matching decoder~\cite{Kubica2022}.
Using different decoders for each stage allows us to investigate which stage limits the overall performance.
For each decoder we estimated the error threshold for $t=4$ QEC cycles, the results are shown in the following table:
\begin{center}
	\renewcommand{\arraystretch}{1.25}
	\begin{tabular}{
		!{\vrule width 1pt}c!{\vrule width 1pt}c!{\vrule width 1pt}c!{\vrule width 1pt}}
		\specialrule{1pt}{0pt}{0pt}
		\rowcolor[gray]{.9}[\tabcolsep] Validation & Correction & Threshold estimate \\
		\specialrule{1pt}{0pt}{0pt}
		Clustering                                 & Clustering & 0.36(1)\%          \\
		Clustering                                 & Matching   & 0.89(2)\%          \\
		Matching                                   & Clustering & 0.38(1)\%          \\
		Matching                                   & Matching   & 1.05(3)\%          \\
		\specialrule{1pt}{0pt}{0pt}
	\end{tabular}
\end{center}
The results suggest that the correction stage is the bottleneck of the decoder, as changing the validation decoder has a comparatively small impact on the performance.
This makes intuitive sense as the flux validation stage uses measurement outcomes that need to satisfy local symmetries captured by the Gauss law, which are not present at the qudit correction stage.
Moreover, this may also explain the fact that the error threshold does not seem to depend on the number of QEC cycles in \cref{fig:threshold}.

Lastly, we emphasize that our implementation of the decoders for the \SAQD{} codes is relatively naive.
For instance, our results could likely be improved by optimizing the clustering decoder using a similar approach as previous work on decoding the two-dimensional qudit toric code~\cite{DuclosCianci2013,DuclosCianci2014,Anwar2014,Watson2015,Hutter2015}.
It may also be possible to improve the performance by adapting the ``single-stage'' decoder described in~\cite{Higgott2023} to the case of subsystem codes.
Another possibility of improving the performance is to view the \SAQD{} codes through the lens of Floquet codes~\cite{Hastings2021}.
In particular, we made an implicit assumption on the measurement schedule to comprise measuring all Pauli $X$-type operators, followed by measuring all Pauli $Z$-type operators.
Considering other (interleaved) measurement schedules is likely to boost error thresholds, as numerically demonstrated with two-dimensional topological codes~\cite{Higgott2021}.


\section{Phases and transitions of an \SAQD{} Hamiltonian model}\label{sec:gapless}

The question as to whether topological subsystem codes in three spatial dimensions are thermally stable is an important open problem~\cite{Bombin2015,Bombin2015a,Brown2016,Burton2018,Bacon2006,Pastawski2010}.
Given a subsystem code, we can associate to it a Hamiltonian, and then a natural question related to its thermal stability is whether the associated Hamiltonian is gapped or gapless.
In particular, the Hamiltonian being gapless would seem to preclude it from being thermally stable.

In this section, we argue that the most natural Hamiltonian associated to the \SAQD{} codes is likely to be at a phase transition.
Our argument follows that of Kramers and Wannier in \onlinecite{Kramers1941}.
In particular, we provide a self-duality on the Hamiltonian, and show that there are at least two phases.
If there is a single transition, it must occur at the point most naturally associated to the code.

Unlike the case of stabilizer codes, associating a Hamiltonian to a subsystem code is not unique.
In this work, we have taken the perspective that the green and yellow AQD codes should form the building blocks of the code.
It is therefore natural to define the code Hamiltonian associated to a group $G$ to be
\begin{align}
	H(J_g,J_y) = & -J_g\!\!\!\!\!\sum_{v,f\in\,\substack{\text{green}  \\\text{spheres}}} (A_v+B_f)\nonumber\\
	             & -J_y\!\!\!\!\!\sum_{v,f\in\,\substack{\text{yellow} \\\text{spheres}}} (A_v+B_f) + h.c.,\label{eqn:codeHamiltonian}
\end{align}
corresponding to \cref{eqn:abelianQDHamiltonian} on the green and yellow spheres with weights $J_g$ and $J_y$ respectively.
Since, from the error correction perspective, there is no preference of either color, the code is associated to $H(1,1)$.

From this perspective, the stabilizer operators in \cref{fig:bulkStabilizers} are (1-form) symmetries of the Hamiltonian.
In \cref{app:gapless}, we show that \cref{eqn:codeHamiltonian} is local-unitary equivalent to a Hamiltonian of the form
\begin{align}
	\tilde{H}(J_g,J_y) & =-J_g\sum_{e\in\text{edges}} X_e -J_y\sum_{e\in\text{edges}} K_e, \label{eqn:SPTHamiltonian}
\end{align}
where
\begin{align}
	\!\!
	K_x=\!\!\!\includegraphicsarray{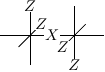}
	,K_y=\!\!\!\includegraphicsarray{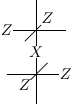}
	,K_z=\!\!\!\includegraphicsarray{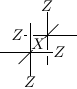}
\end{align}
are reminiscent of the Raussendorf-Bravyi-Harrington (RBH) model~\cite{Raussendorf2005}.

\cref{eqn:SPTHamiltonian} has a self-duality map (see \cref{app:gapless}) exchanging $J_g\leftrightarrow J_y$.
When $(J_g,J_y)=(1,0)$, the Hamiltonian is paramagnetic.
Conversely, at the point $(0,1)$, there is non-trivial 1-form $(\ZZ{2}\times\ZZ{2})$-protected topological order, where the symmetries are generated by the images of the stabilizers.

From this self-duality, we can deduce that if the model has a single phase transition, then it must occur where $J_g=J_y$.
If this transition is continuous, it follows that $H(1,1)$ is gapless.
It could also be that the transition is first order, as in the symmetric sector~\cite{Li2023}, in which case the gap would remain open.

It has previously been noted that thermal stability of the RBH model with 1-form symmetries may be closely related to the single-shot properties of associated subsystem codes~\cite{Roberts2017,Kubica2022}.
Our perspective suggests that the relevant symmetries are the stabilizer operators of the subsystem code.


\section{Discussion}\label{sec:remarks}

In this article, we provided a systematic construction of the three-dimensional \SAQD{} codes, which provide a natural generalization of the \STC{}.
Although we focused on the cubic lattice, the \SAQD{} codes can be realized on colorable octahedral lattices~\cite{Kubica2022}.
It is currently unclear which aspects of these lattices are necessary and which are merely sufficient.

Our construction required inflating the vertices of the lattice to 2-spheres on which the two-dimensional topologically-ordered model was placed.
The choice of spheres does not seem to be essential, rather one could consider inflating to more interesting 2-manifolds.
In such a setting, the constraints on the gauge flux described in \cref{sec:errorConfigs} should remain in place, however other code properties may be more exotic.
In order to ensure that the \SAQD{} codes exhibit the single-shot QEC property, we had to carefully choose the geometry used to combine these 2-spheres.
Manifolds with non-trivial genus would allow exploration of more interesting geometries, such as linking.

Central to our construction were topologically-ordered models in two spatial dimensions, which we assumed to be untwisted abelian quantum double models.
It is natural to ask whether our construction can be extended to all abelian anyon models, or general Levin-Wen models.
From a physical viewpoint, such a generalization may lead to novel phases, possibly without a gap.
This raises the question of classifying the \SAQD{} codes and other topological subsystem codes in three spatial dimensions, and whether there is a connection with the classification of two-dimensional topological orders.
To properly formulate this question, we require a notion of equivalence between (topological) subsystem codes, which generalizes local unitary equivalence of topological subspace codes~\cite{Ellison2023}.
Beyond pure-state properties, the authors of \onlinecite{Zini2021} have defined a notion of mixed-state topological quantum field theory (TQFT), and established a connection to the two-dimensional Levin-Wen models based on modular fusion categories.
It would be interesting to explore this connection, and understand whether the \SAQD{} codes give more examples of mixed-state TQFTs.

The gauge color code appears to fit into our perspective, but not directly into our construction.
Around each volume of the \GCC{}, the gauge generators form a copy of the stabilizer color code model in two spatial dimensions.
This is reminiscent of the AQD models around vertices in our construction, and would suggest the gauge color code should be equivalent to the \SAQD{} associated to $\ZZ{2}\times\ZZ{2}$.
This appears to be in slight conflict with the stabilizer case.
In three spatial dimensions, the stabilizer color code (without boundaries) is locally-unitarily equivalent to three decoupled copies of the stabilizer toric code~\cite{Kubica2015}.
We suspect that resolving this disparity would shed light on both the appropriate notion of equivalence and the necessary three-dimensional geometrical properties.

The \SAQD{} codes constitute an excellent illustration of the single-shot QEC property and physics behind it.
Moreover, they can provide illuminating insights into a connection between single-shot QEC and self-correcting memories.
We have argued that the most naive Hamiltonians associated to the \SAQD{} codes are probably gapless, but are closely related to models with symmetry-protected self-correction~\cite{Roberts2020,Kubica2018,Stahl2021}.
Given this close connection, one might ask whether the subsystem code Hamiltonian could be modified to energetically enforce the symmetry, possibly leading to a self-correcting quantum memory.


\acknowledgments
J.B.~was funded by the Research Foundation\textendash Flanders via grant G087918N and postdoctoral fellowship 12E9223N.
Research at Perimeter Institute is supported in part by the Government of Canada through the Department of Innovation, Science and Industry Canada and by the Province of Ontario through the Ministry of Colleges and Universities.
This project (EOS 40007526) has received funding from the FWO and F.R.S.-FNRS under the Excellence of Science (EOS) programme.
We would like to thank Hussain Anwar, Lander Burgelman, Elijah Durso-Sabina, Pei Jiang, and Sam Roberts for helpful discussions.
We thank Lander Burgelman for comments on an earlier draft.


\emph{Note added.---}We would like to bring the reader’s attention to a related, independent work by Li, von Keyserlingk, Zhu, and Jochym-O’Connor~\cite{Li2023}, which explores a phase diagram of the subsystem toric code and appeared in the same arXiv posting.


%


\onecolumngrid
\appendix
%

\section{Pauli subsystem codes}\label{app:pauliSubsystemCodes}

For completeness, in this section we will review the definition of a Pauli subsystem code following \onlinecite{Poulin2005,Bombin2010a}.

Setting notation:
Let $G$ be a group.
\begin{subalign}
	\text{Center: }     &  & \cent{G}:=  & \set{g\in G}{ghg^{-1}=h,\,\forall h\in G},                            \\
	\text{Rank: }       &  & \rk{G}:=    & \text{size of minimal generating set},                                \\
	\text{Order: }      &  & \order{G}:= & \text{number of elements of $G$},                                     \\
	\text{Commutator: } &  & [g,h]:=     & ghg^{-1}h^{-1}                             & \text{for $g,\,h\in G$}.
\end{subalign}

We begin with an abstract definition of the Pauli group for $n$ qudits of dimension $q$.
Later we will act on the vector space $(\mathbb{C}^{q})^{\otimes n}$ using the standard representation.

\begin{definition}[Pauli group]
	Let $n$ and $q$ be positive integers.
	The \define{$(q,n)$-Pauli group} is defined by
	\begin{align}
		\Pauli{q}{n}:= &
		\begin{cases}
			\grouppresentation{\alpha_{q},X_1,\ldots,X_n,Z_1,\ldots,Z_n}{X_i^q=Z_i^q=\alpha_{q}^{2q}=\alpha_{q}^{2}\comm{X_i,Z_i}=1} & \text{$q$ even} \\
			\grouppresentation{\alpha_{q},X_1,\ldots,X_n,Z_1,\ldots,X_n}{X_i^q=Z_i^q=\alpha_{q}^q=\alpha_{q}\comm{X_i,Z_i}=1}        & \text{$q$ odd.}
		\end{cases}
	\end{align}
	The group has order and rank
	\begin{subalign}
		\order{\Pauli{q}{n}}= &
		\begin{cases}
			2q^{2n+1} & \text{$q$ even} \\
			q^{2n+1}  & \text{$q$ odd},
		\end{cases} \\
		\rk{\Pauli{q}{n}} =   &
		\begin{cases}
			2n+1 & \text{$q$ even} \\
			2n   & \text{$q$ odd}.
		\end{cases}
	\end{subalign}
	Each element of $\Pauli{q}{n}$ can be expressed uniquely as
	\begin{align}
		g & =\alpha_{q}^{a}X_1^{x_1}X_2^{x_2}\ldots X_n^{x_n}Z_1^{z_1}Z_2^{z_2}\ldots Z_n^{z_n},\label{eqn:pauliGroupWord}
	\end{align}
	for some $0\leq x_i,z_i<q$, and
	\begin{align}
		0\leq a<\begin{cases}
			        2q & \text{ $q$ even} \\
			        q  & \text{ $q$ odd}.
		        \end{cases}
	\end{align}
	The center of $\Pauli{q}{n}$ is
	\begin{align}
		\cent{\Pauli{q}{n}}= & \grouppresentation{\alpha_{q}}{}\cong
		\begin{cases}
			\ZZ{2q} & \text{$q$ even} \\
			\ZZ{q}  & \text{$q$ odd.}
		\end{cases}
	\end{align}
\end{definition}

\begin{definition}[Stabilizer group]
	A \define{stabilizer group} is an abelian subgroup $\Stab\subgp\Pauli{q}{n}$ such that
	\begin{align}
		\Stab\cap\cent{\Pauli{}{}}= & \{1\}.
	\end{align}
	Following \onlinecite{Bombin2010a}, a stabilizer group can be specified by supplying a pair $(\phi, s)$, where $\phi$ is an automorphism of $\Pauli{q}{n}$, and $s\leq n$ is the rank of $\Stab$.
	The $s$ generators of $\Stab$ are the images of $X_1,\ldots,X_s\in\Pauli{q}{n}$
	\begin{align}
		\Stab:= & \grouppresentation{\phi(X_1),\phi(X_2),\ldots,\phi(X_s)}{}.
	\end{align}
	The group has order and rank
	\begin{subalign}
		\order{\Stab}= & q^s, \\
		\rk{\Stab} =   & s.
	\end{subalign}
	We remark that there is a large redundancy in this description, with many choices of $\phi$ leading to the same $\Stab$.
\end{definition}

\begin{definition}[Gauge group]
	A \define{gauge group} $\G\subgp\Pauli{q}{n}$ is specified by a triple $(\phi,s,r)$, where $\phi$ is an automorphism of $\Pauli{q}{n}$ and the non-negative $s$, $r$ obey $s+r\leq n$.
	Provided with this data, the gauge group is defined as
	\begin{subalign}
		\G:= & \grouppresentation{\alpha_{q},\phi(X_1),\phi(X_2),\ldots,\phi(X_s),\phi(X_{s+1},\ldots,\phi(X_{s+r}),\phi(Z_{s+1}),\ldots,\phi(Z_{s+r})}{} \\
		=    & \grouppresentation{\alpha_{q},\Stab,\phi(X_{s+1},\ldots,\phi(X_{s+r}),\phi(Z_{s+1}),\ldots,\phi(Z_{s+r})}{},
	\end{subalign}
	where $\Stab = \grouppresentation{\phi(X_1),\phi(X_2),\ldots,\phi(X_s)}{}$ is the stabilizer subgroup.

	The group $\G$ has order and rank
	\begin{subalign}
		\order{\G}= &
		\begin{cases}
			2q^{s+2r+1} & \text{$q$ even} \\
			q^{s+2r+1}  & \text{$q$ odd},
		\end{cases} \\
		\rk{\G} =   &
		\begin{cases}
			s+2r+1 & \text{$q$ even} \\
			s+2r   & \text{$q$ odd}.
		\end{cases}
	\end{subalign}
\end{definition}

\begin{definition}[Bare logical group]
	Given a gauge group $\G$ defined by the triple $(\phi,s,r)$, the \define{bare logical group} is
	\begin{align}
		\bareLog & = \grouppresentation{\alpha_{q},\phi(X_{s+r+1}),\phi(Z_{s+r+1}),\phi(X_{s+r+2}),\phi(Z_{s+r+2}),\ldots,\phi(X_{s+r+k}),\phi(Z_{s+r+k})}{},
	\end{align}
	where $s+r+k=n$.
	The minimal length of a non-trivial (not only containing powers of $\alpha_{q}$) word $W$ in $\bareLog$ is the \define{bare distance} $\bareDistance$.
	Equivalently, $W$ is non-trivial if $\phi^{-1}(W)$ contains at least one element of $\{X_{s+r+1},Z_{s+r+1},X_{s+r+2},Z_{s+r+2},\ldots,X_{s+r+k},Z_{s+r+k}\}$ when written in standard form \cref{eqn:pauliGroupWord}.
\end{definition}

\begin{definition}[Dressed logical group]
	Given a gauge group $\G$ defined by the triple $(\phi,s,r)$, the dressed logical group is
	\begin{align}
		\dressedLog = \grouppresentation{\alpha_{q},\G,\phi(X_{s+r+1}),\phi(Z_{s+r+1}),\phi(X_{s+r+2}),\phi(Z_{s+r+2}),\ldots,\phi(X_{s+r+k}),\phi(Z_{s+r+k})}{}.
	\end{align}
	The minimal length of a non-trivial word in $\dressedLog$ is the \define{dressed distance} $\dressedDistance$.
\end{definition}

\begin{definition}[Pauli subsystem code]
	Given a gauge group $\G$ acting on $n$ physical qudits by the standard representation, the stabilizer group partitions the Hilbert space into two subspaces
	\begin{align}
		(\mathbb{C}^q)^n\cong \C\oplus\C^{\perp},
	\end{align}
	where $\C$ is the subspace on which $\Stab$ acts as $+1$, with dimension $q^{n-s}$.
	In a Pauli subsystem code, the code space $\C$ is further decomposed into
	\begin{align}
		\C\cong \mathcal{A}\otimes\mathcal{B},
	\end{align}
	where $\mathcal{A}$ is the logical subsystem, and $\mathcal{B}$ is called the gauge space.
	The gauge group $\G$ acts trivially on the $q^k$-dimensional subsystem $\mathcal{A}$, while $\bareLog\cong\Pauli{q}{k}$ acts as the logical group on this system.
	The remaining $q^r$-dimensional subsystem $\mathcal{B}$ has non-trivial $\G$ action.

	Pragmatically, the gauge space is a `junk' space that can be used to bypass no-go theorems for pure stabilizer codes.

	A Pauli subsystem code on $n$ physical qudits, each of dimension $q$, which encodes $k$ logical qudits with bare distance $\bareDistance$ and dressed distance $\dressedDistance$ has parameters
	\begin{align}
		\codeparameters{n,k,\bareDistance,\dressedDistance}{q}.
	\end{align}
	It is often convenient to express the various groups in terms of a generating set.
	The number of encoded qudits $k$ can be calculated from the groups by
	\begin{align}
		k = & n-(s+r) \\
		=   &
		\begin{cases}
			n-\frac{\rk{\G}+\rk{\S}-1}{2} & \text{$q$ even} \\
			n-\frac{\rk{\G}+\rk{\S}}{2}   & \text{$q$ odd}.
		\end{cases}
	\end{align}

	Typically, a Pauli subsystem code is specified by providing an (over-complete) generating set for the gauge group $\G$, usually omitting mention of $\alpha_{q}$.
	The stabilizer subgroup then consists of the center of the group defined by the given generators, ignoring phases.
	We follow this tradition in the main text.
\end{definition}


\section{Operators for \texorpdfstring{$\ZZ{d}$}{Zd} subsystem quantum double code with various boundary conditions}\label{app:explicitOperators}

In this appendix, we provide a complete list of gauge, stabilizer, and logical operators for the $\ZZ{d}$ code with various choices of boundary conditions.
Recall that the model is defined on a cubic lattice, with qudits on edges.
Each vertex is surrounded by a tetrahedron embedded on the surface of an oriented sphere.
The specific code instance, although not the code properties, depends on a choice of orientation for the edges of the tetrahedra.
For concreteness, we choose the translationally invariant orientations
\begin{align}
	\includegraphicsarray{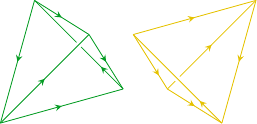}.\label{eqn:edgeDirections_Appendix}
\end{align}
For simplicity, we will assume in all cases that the cubic lattice has dimensions $L\times L\times L$, where $L$ is even.

\href{https://mikevasmer.github.io/qudit-single-shot/}{To aid visualization, we provide interactive figures corresponding to many of the discussions below at }\onlinecite{website}.

\subsection{Bulk/\texorpdfstring{$\Torus{3}$}{T3}}\label{sec:bulkOperators}

Recall that the gauge operators of the \SAQD{} code correspond to the stabilizer operators (\cref{eqn:abelianQDTerms}) of $\ZZ{d}$ quantum double models defined on these tetrahedra.
Local generators for the gauge group are therefore given by
\begin{subalign}
	\includegraphicsarray{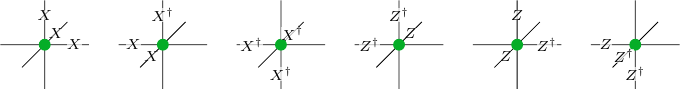}\label{eqn:bulkGreenGaugeOps}
	\\
	\includegraphicsarray{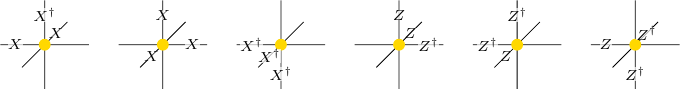},\label{eqn:bulkYellowGaugeOps}
\end{subalign}
where $X^\dagger=X^{-1}$ and $Z^\dagger=Z^{-1}$, and the vertex color indicates the color of the enclosing sphere.

\begin{figure}\centering
	\includegraphicsarray{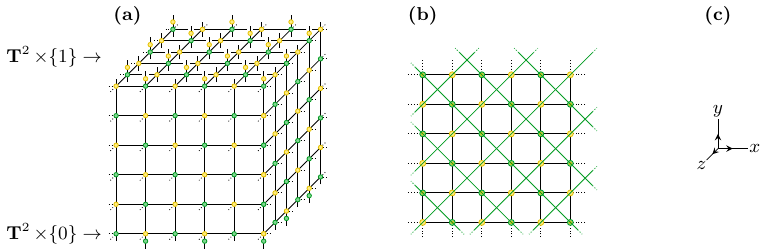}
	{\phantomsubcaption\label{fig:openTopLattice3D}}
	{\phantomsubcaption\label{fig:openTopLattice2D}}
	{\phantomsubcaption\label{fig:appendixCoordinates}}
	\caption{$\Torus{2}\times\Interval$, 2-dimensional AQD models are on the top and bottom boundaries, with lattice periodic in other directions.
		We regard the `dangling' edges as having length $1/2$, meaning this lattice has linear size $6\times 6\times 6$.
			{\bf(b)} shows the view from above.
		All spheres incident on the top boundary are colored yellow, and the AQD model on the surface is colored green.
		Colors are reversed at the bottom boundary.}\label{fig:openTopLattice}
\end{figure}

Stabilizer operators in the bulk correspond to volumes in the lattice.
Volumes that only contain faces of tetrahedra give $X$ type stabilizers, whilst volumes containing vertices correspond to $Z$ type stabilizers.
Each stabilizer can be recovered from an entirely green set of gauge generators, and also an entirely yellow set:
\begin{subalign}[eqn:bulkStabilizers]
	\includegraphicsarray{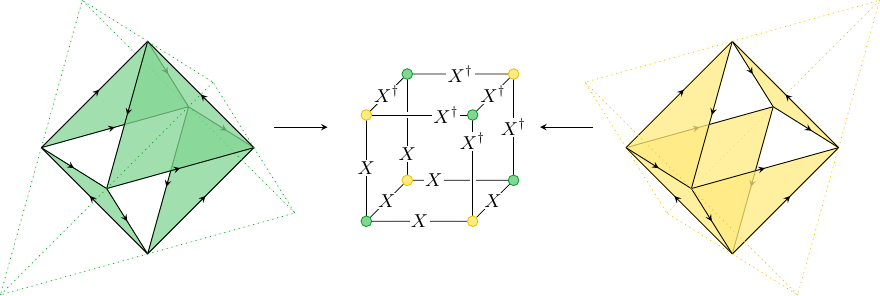}\label{eqn:Xstablilizer}
	\\
	\includegraphicsarray{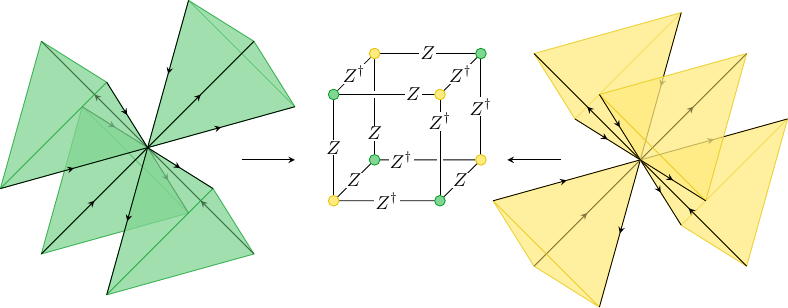}.\label{eqn:Zstabilizer}
\end{subalign}
In addition to these `locally generated' stabilizers, there are stabilizers corresponding to the 3 coordinate planes (\cref{fig:appendixCoordinates}), for example the $xy$-stabilizer generators are
\begin{subalign}[eqn:sheetStabilizers]
	\includegraphicsarray{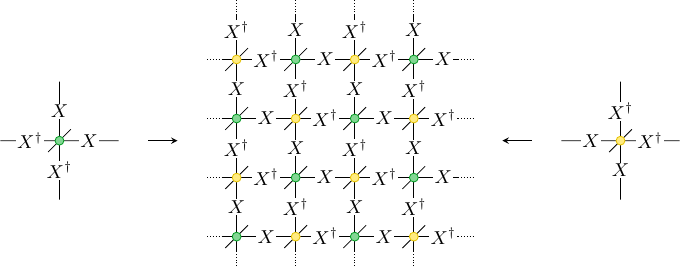}
	\\
	\includegraphicsarray{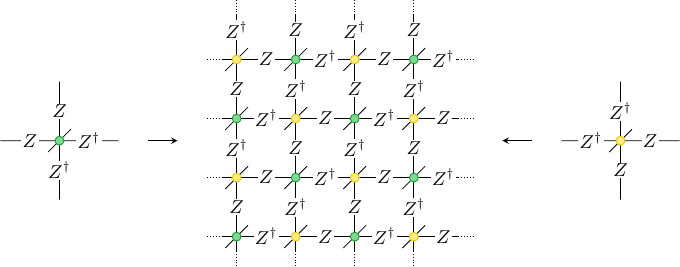}.
\end{subalign}

On the cubic lattice of dimension $L$, there are
\begin{align}
	n=3L^3
\end{align}
qudits.
Each volume contributes a single local stabilizer generator.
The product of all $X$ type local stabilizer generators is the identity operator, with the same being true of the $Z$ type (local) stabilizer generators.
We therefore find that
\begin{align}
	\rk\S=(L^3-2)+2 \times 3=L^3+4,
\end{align}
where the $-2$ comes from the relation on the local generators, and the $+2\times 3$ is contributed by the sheet-like generators.

There are 6 gauge generators associated to each vertex, however there are some constraints.
In particular, there are two ways to realize each stabilizer from gauge generators, so one constraint is contributed for each stabilizer generator.
We therefore find that
\begin{align}
	\rk\G & =
	\begin{cases}
		6 L^3-\rk\S+1 & d\text{ even} \\
		6 L^3-\rk\S   & d\text{ odd}  \\
	\end{cases}.
\end{align}
Finally, we find that there are $k=0$ encoded qudits on on $\Torus{3}$.

\subsection{\texorpdfstring{$\Torus{2}\times \I$}{T2xI}}\label{sec:T2xI}

We now consider partially open boundary conditions.
In the bulk, nothing is changed from \cref{sec:bulkOperators}.
At the boundaries $\Torus{2}\times\{0\}$ and $\Torus{2}\times\{1\}$, we place 2-dimensional AQD models of opposite colors.
To achieve this, we need to modify the lattice at the boundary, adding qudits on `half-edges' as indicated on the top surface of \cref{fig:openTopLattice}.
In that figure, new spheres have been attached to the top (all yellow) and bottom (all green).
Unlike those in the bulk, these are only in contact with two lattice edges, so are only decorated with two oriented edges.
We choose the following everywhere for simplicity
\begin{align}
	\includegraphicsarray{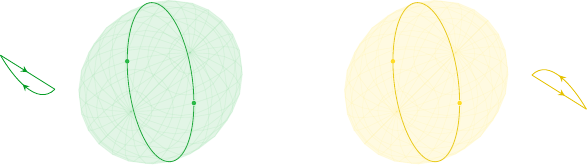},
\end{align}
where we remark that the orientation bottom edge of the yellow spheres is dictated by the choice on bulk spheres, and similarly for the top edge of the green spheres.
Inclusion of these additional spheres, and their AQD models, leads to the additional gauge generators
\begin{align}
	\includegraphicsarray{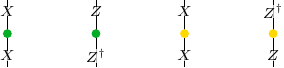}.
\end{align}

With these extra qudits in place, we can assign AQD operators to the top and bottom faces.
These are colored with the opposite color to the spheres incident on the face
\begin{align}
	\includegraphicsarray{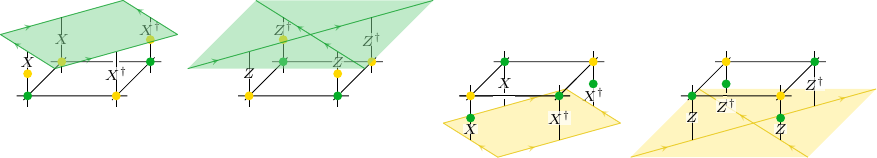},
\end{align}
completing the gauge generators for the model on $\Torus{2}\times\Interval$.

Compared to $\Torus{3}$, there are $L^2$ additional qudits contributed by not identifying the top and bottom edges.
Additionally, there are $L^2/2$ extra qudits introduced on both the bottom and top faces.
Therefore, there are
\begin{align}
	n=3L^3+2L^2
\end{align}
qudits on the lattice.

The bulk stabilizers from \cref{eqn:bulkStabilizers} remain stabilizers for these boundary conditions, however the sheet-like stabilizers of \cref{eqn:sheetStabilizers} can no longer be constructed from gauge operators in the $xy$ and $yz$-planes.
As we explain below, they become bare logical operators.
In the vicinity of the boundary, we obtain new locally generated stabilizers
\begin{subalign}[eqn:topbottomStabilizers]
	\includegraphicsarray{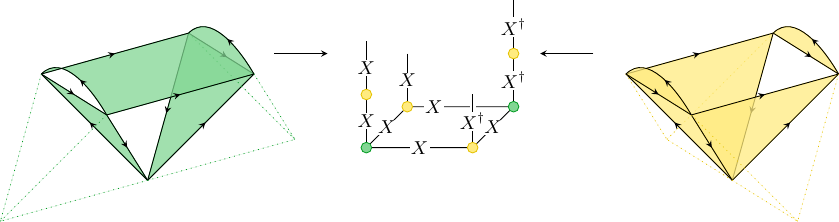}\label{eqn:XstablilizerTop}
	\\
	\includegraphicsarray{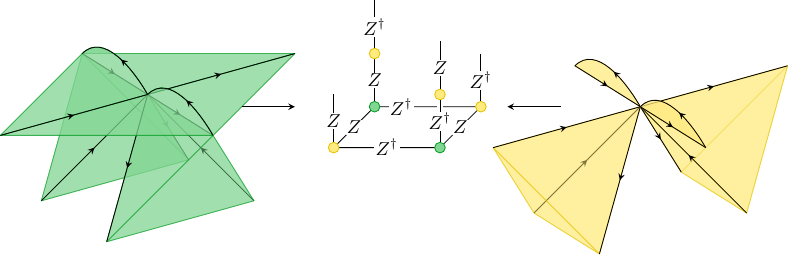}\label{eqn:ZstabilizerTop}
	\\
	\includegraphicsarray{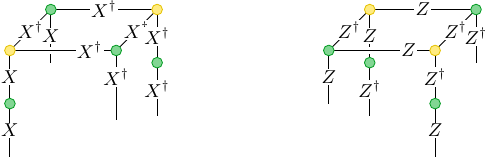}.\label{eqn:stabilizerBottom}
\end{subalign}

Each bulk locally generated stabilizer remains a stabilizer, in addition to the `chopped' stabilizers at the boundary.
The $xy$ and $yz$ sheet-like operators are no longer stabilizers, and the $xz$ sheet can be produced from the boundary operators \cref{eqn:topbottomStabilizers}.
We therefore find that
\begin{align}
	\rk \S & =2\left(\frac{L^3}{2}+\frac{L^2}{2}-1\right),
\end{align}
where the $-1$ arises because the product of all $X$- or $Z$-type stabilizers is the identity.
Finally, there are the same number of gauge generators in the bulk.
Additionally, each new sphere on the two boundaries contributes two generators.
The AQD models on the two faces contribute $L^2-2$ gauge generators each, where the $-1$ occurs because the product of all these generators is identity for both the $X$ and $Z$ types.
As in the bulk, each stabilizer generator can be produced in two ways, adding $\rk\S$ constraints.
We therefore find
\begin{align}
	\rk \G
	 & =\left\{
	\begin{array}{lll}
		6 L^3+\frac{2\ 2 L^2}{2}+2 \left(L^2-2\right) - \rk \S+1 & =5 L^3+3 L^2-1 & d\text{ even} \\
		6 L^3+\frac{2\ 2 L^2}{2}+2 \left(L^2-2\right) - \rk \S   & =5 L^3+3 L^2-2 & d\text{ odd}
	\end{array}\right. .
\end{align}
Finally, we have
\begin{align}
	k=3L^3+2L^2-\frac{5 L^3+3 L^2-2+L^3+L^2-2}{2}=2
\end{align}
encoded qudits.

\subsubsection{Logical operators}

\begin{figure}\centering
	\includegraphicsarray{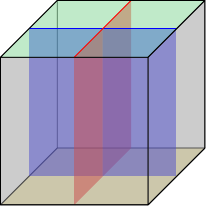}
	\caption{
		On $\Torus{2}\times\Interval$, the code supports two qudits.
		Bare logical operators are given by sheets of $X$ and $Z$ operators in the $xy$ and $yz$ planes.
		Minimal weight dressed logical operators are given by the intersection of the plane with the upper and lower faces.
		Here we indicate one logical pair. \interactivefig}\label{fig:T2xILogicals}
\end{figure}

On $\Torus{3}$, there are sheet-like stabilizer operators.
These can no longer be built from gauge generators on $\Torus{2}\times\Interval$, but continue to commute with the gauge group, and so are promoted to bare logical operators.
There are two pairs of bare logical operators, given by $X$ or $Z$ sheets in the $xy$ and $yz$ planes.
Example pairs are
\begin{align}
	\includegraphicsarray{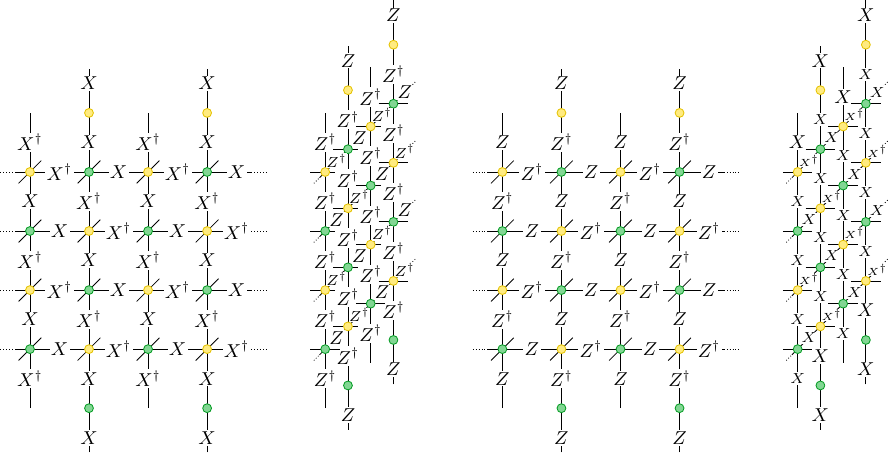}.\label{eqn:sheetLogicals}
\end{align}
The associated (minimum weight) dressed logical operators can be obtained by using the gauge operators in \cref{eqn:sheetStabilizers} to reduce the sheet-like bare logical to string-like operators on the AQD models at $\Torus{2}\times\{0\}$ or $\Torus{2}\times\{1\}$.
It can readily be verified that these operators have weight $\bareDistance=2 L (L+1)$ and $\dressedDistance=L$ respectively.

Finally, the $\ZZ{d}$ code on $\Torus{2}\times\Interval$ has code parameters
\begin{align}
	\codeparameters{n,k,\bareDistance,\dressedDistance}{d} & =\codeparameters{3L^3+2L^2,2,2 L (L+1),L}{d}.
\end{align}

\subsection{\texorpdfstring{$\Interval^3$}{IxIxI}}\label{app:cubeOperators}

\begin{figure}\centering
	\includegraphicsarray{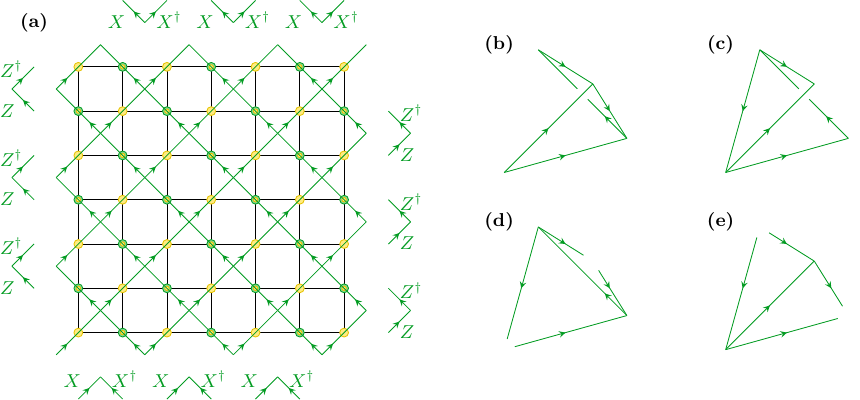}
	{\phantomsubcaption\label{fig:boundaryGaugeA}}
	{\phantomsubcaption\label{fig:boundaryGaugeB}}
	{\phantomsubcaption\label{fig:boundaryGaugeC}}
	{\phantomsubcaption\label{fig:boundaryGaugeD}}
	{\phantomsubcaption\label{fig:boundaryGaugeE}}
	\caption{%
		Boundary (green) gauge generators.
		Left and right faces are smooth, front and back are rough.
			{\bf(a)} shows a top view, {\bf (b--e)} show the green tetrahedra incident on the boundaries.
		At the left/right boundaries, partial vertex terms (2-body) are included, but partial faces are not.
		At the front/back boundaries, partial faces are included, but partial vertex terms are omitted.
		Yellow gauge generators are analogous.}\label{fig:boundaryGauge}
\end{figure}

We now consider the model on a cube.
This is similar to the case in \cref{sec:T2xI}, but the boundaries $M\times\{0\}$ and $M\times\{1\}$ are changed to AQD models, themselves with boundaries.
Specifically, we place AQD models with two rough and two smooth boundaries on these faces.
As (2+1)-dimensional stabilizer models, these support a single qudit.
Despite incorporating two such codes, the \SAQD{} codes also encodes a single qudit, with the additional dimension giving the enhanced protection of single-shot error correction.

In the bulk, and the interior of the top and bottom boundaries, the gauge generators remain the same as in \cref{sec:T2xI}.
We will identify the left/right faces as smooth, and the front/back boundaries as rough.
Any two-dimensional AQD model incident on a rough/smooth boundary is given rough/smooth boundary conditions, with all terms included as gauge operators rather than stabilizers.
The top AQD model, and the green spheres are shown in \cref{fig:boundaryGauge}, with the bottom and yellow spheres being similar.

Finally, at the intersection of faces, spheres with multiple types of boundary conditions (appropriate to the incident face) are chosen.
The complete list of gauge generators on $\Interval^3$ is, in addition to those in \cref{sec:T2xI},
\begin{outline}
	\1 Left:
	\begin{align}
		\includegraphicsarray{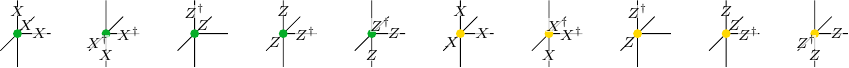}
	\end{align}
	\1 Right:
	\begin{align}
		\includegraphicsarray{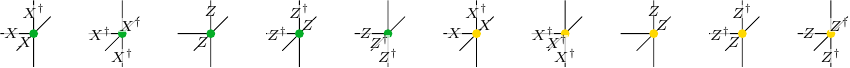}
	\end{align}
	\1 Back:
	\begin{align}
		\includegraphicsarray{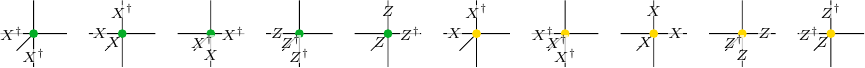}
	\end{align}
	\1 Front:
	\begin{align}
		\includegraphicsarray{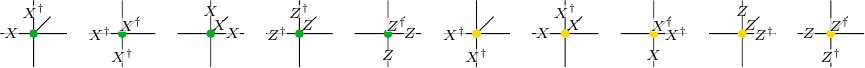}
	\end{align}
	\1 Front left:
	\begin{align}
		\includegraphicsarray{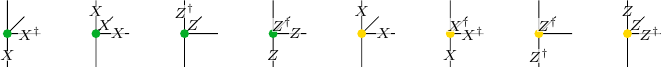}
	\end{align}
	\1 Front right:
	\begin{align}
		\includegraphicsarray{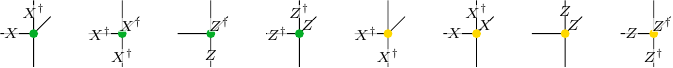}
	\end{align}
	\1 Back left:
	\begin{align}
		\includegraphicsarray{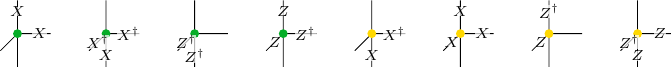}
	\end{align}
	\1 Back right:
	\begin{align}
		\includegraphicsarray{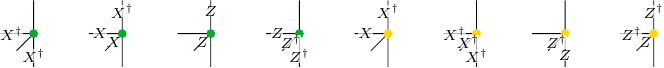}
	\end{align}
	\1 Top boundaries:
	\begin{align}
		\includegraphicsarray{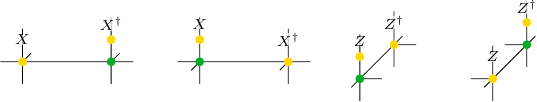}
	\end{align}
	\1 Bottom boundaries:
	\begin{align}
		\includegraphicsarray{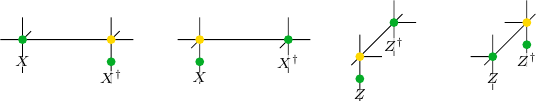}
	\end{align}
\end{outline}

\begin{figure}\centering
	\includegraphicsarray{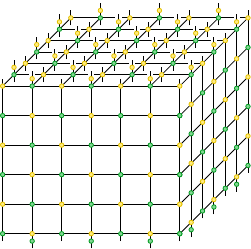}
	\caption{%
		Example lattice on $\Interval^3$.
		We regard the `dangling' edges as having length $1/2$, meaning this lattice has linear size $6\times 6\times 6$.}\label{fig:latticeIxIxI}
\end{figure}

The code on $\Interval^3$ is very similar to the one on $\Torus{2}\times\Interval$, however the lattice geometry causes the counting of qudits to be slightly altered.
Restricting our attention to $L^3$ lattices, with $L$ even, for example that depicted in \cref{fig:latticeIxIxI}, there are
\begin{align}
	n=3L^3+6L^2+5L+1
\end{align}
qudits.

All bulk and top/bottom stabilizers are carried over from \cref{sec:T2xI}.
In addition, there are stabilizers associated to the newly introduced faces.
As before, these are operators that can be constructed as the product of gauge generators of either color.
These operators are
\begin{align}
	\includegraphicsarray{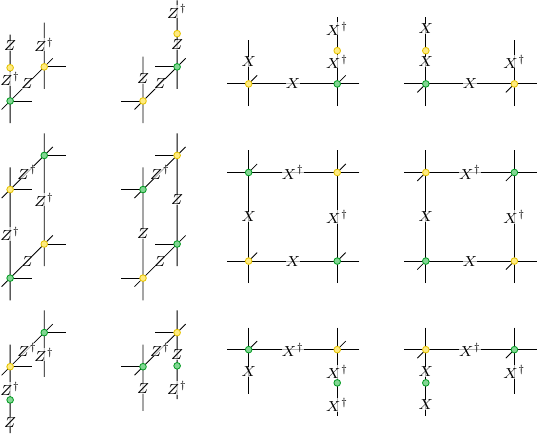}.
\end{align}

In total, there are $L^2(L-1)$ stabilizer generators associated to the bulk, $L^2$ to both the bottom and top, and $L(L+1)/2$ to each of the remaining faces.
Unlike the periodic lattices, there are no relations among these generators, so
\begin{align}
	\rk\S & =L^3+3 L^2+2 L.
\end{align}
Accounting for all the local generators, there are $2\times(3 L^3+6 L^2+5 L)$.
Each stabilizer generator introduces a relation, so we find
\begin{align}
	\rk \G
	 & =\left\{
	\begin{array}{lll}
		6 L^3+12 L^2+10 L - \rk \S+1 & =5 L^3+9 L^2+8 L+1 & d\text{ even} \\
		6 L^3+12 L^2+10 L - \rk \S   & =5 L^3+9 L^2+8 L   & d\text{ odd}
	\end{array}\right.,
\end{align}
and
\begin{align}
	k & =3L^3+6L^2+5L+1-\frac{6 L^3+12 L^2+10 L}{2}=1.
\end{align}

\subsubsection{Logical operators}

The logical operators of this code are inherited directly from those in \cref{sec:T2xI}.
The first pair from \cref{eqn:sheetLogicals} no longer commute with the boundary terms, but the second pair remain as bare logical operators.
The minimum weight dressed logical operators are string operators for the AQD codes on the top or bottom face as in \cref{fig:T2xILogicals}.

Due to the altered geometry, the distance of these operators is modified slightly.
The $\ZZ{d}$ code on $\Interval^3$ has code parameters
\begin{align}
	\codeparameters{n,k,\bareDistance,\dressedDistance}{d} & =\codeparameters{3L^3+6L^2+5L+1,1,2 L^2+3 L+1,L+1}{d}.
\end{align}

\subsection{\texorpdfstring{$\left(\Torus{2}\times \Interval\right)^\prime$}{(T2xI)'}}\label{sec:threeBodyOnly}

Finally, we return to the boundary conditions $\Torus{2}\times \Interval$, and, following \onlinecite{Kubica2022}, describe a way to modify the AQD models on the top and bottom faces to produce a code with all gauge operators having weight 3 or less.
By will introduce additional qudits on the faces and apply a local circuit to reduce the weight of the AQD model gauge terms.
The surface model we obtain is identical to the one introduced in \onlinecite{Bravyi2013} in the case $d=2$, so provides a generalization of the model introduced there to $\ZZ{d}$ AQD models.

Our circuit is formed from generalized controlled-$X$ gates, defined by
\begin{align}
	C_X:=\includegraphicsarray{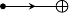}
	 & =\frac{1}{d}\sum_{ij}\omega^{ij}Z^i\otimes X^j
	=
	\left(\includegraphicsarray{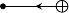}\right)^\dagger,
\end{align}
which act on the $\ZZ{d}$ Pauli operators as
\begin{align}
	C_X XI C_X^\dagger & =XX^\dagger &  & C_X IZ C_X^\dagger  =ZZ.
\end{align}
and reduces to the CNOT gate when $d=2$.
Additional qudits are placed on half of the plaquettes on the top and the bottom faces, for example on the top face
\begin{align}
	\includegraphicsarray{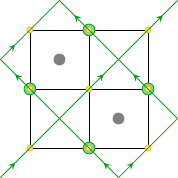},\label{eqn:plaquetteQudits}
\end{align}
where the gray dots indicate the added qudits, which we refer to as plaquette qudits.
For each of the $L^2$ plaquette qudits, we add two gauge generators, $X$ and $Z$.
The transformed code therefore has
\begin{subalign}
	n & =3L^3+3L^2                                                                                 \\
	\rk \G
	& =\left\{
	\begin{array}{lll}
		6 L^3+\frac{2\ 2 L^2}{2}+2 \left(L^2-2\right) - \rk \S+1+2L^2 & =5 L^3+5 L^2-1 & d\text{ even} \\
		6 L^3+\frac{2\ 2 L^2}{2}+2 \left(L^2-2\right) - \rk \S +2L^2  & =5 L^3+5 L^2-2 & d\text{ odd}
	\end{array}\right. \\
	k & =2.
\end{subalign}

The circuit is translationally invariant, acting on a unit cell as
\begin{align}
	\includegraphicsarray{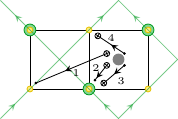},\label{eqn:threeBodyCircuit}
\end{align}
where the gates are applied in the order indicated, with all `1' operators applied first, followed by all `2' operators, and so on.
Note that the gates act only on the qudits sticking out of the surface, and therefore any gauge operator that does not have support on these sites is left unchanged.
The altered gauge generators on the top are
\begin{align}
	\includegraphicsarray{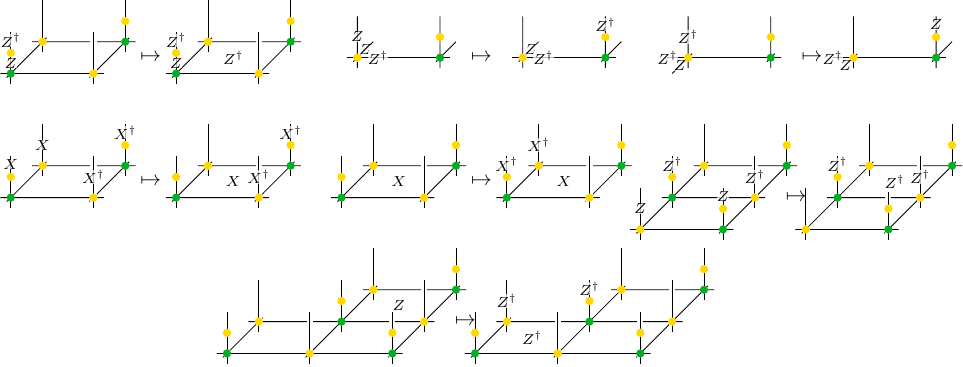},
\end{align}
and on the bottom
\begin{align}
	\includegraphicsarray{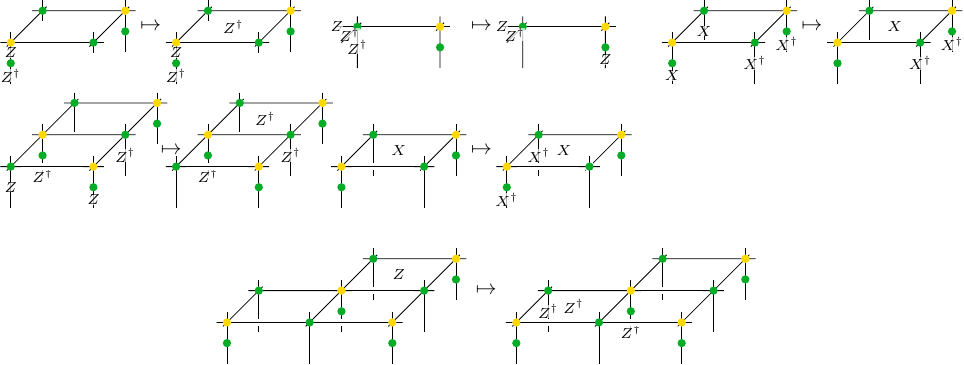}.
\end{align}

Inclusion of the additional qudits alters the code parameters slightly.
We distinguish between the $xy$ and $yz$ qudits, where the label denotes the plane of the bare logical $X$ operator on that qudit.
The code has parameters
\begin{align}
	\codeparameters{n,k,\bareDistance[(xy)],\dressedDistance[(xy)]\middle|\bareDistance[(yz)],\dressedDistance[(yz)]}{d} & =\codeparameters{3L^2(L+1),2,2L(L+1),L\middle|2L(L+2),\frac{L}{2}}{d}.
\end{align}
The logical operators for the $xy$-logical qudit are those of \cref{eqn:sheetLogicals}, while the $yz$-qudit logical operators are deformed by the circuit \cref{eqn:threeBodyCircuit}.
In particular, the dressed logical operators are supported on the line of plaquette qudits (\cref{eqn:plaquetteQudits}) parallel to the associated sheet operator.


\section{Phases and transitions of an \SAQD{} Hamiltonian model}\label{app:gapless}

In this appendix, we argue that the naive Hamiltonians for the codes presented are probably likely to be at a phase transition.
We restrict our attention to the $\ZZ{2}$ case, but the results generalize to all abelian groups.

We can find duality of a single copy of the $\ZZ{2}$ model to a different model with symmetry protected order.
We restrict attention to the 3-torus for simplicity.
Apply the following circuit around each green vertex
\begin{align}
	\includegraphicsarray{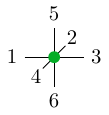}
	 &  &
	\includegraphicsarray{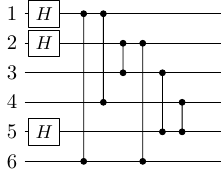}.\label{eqn:toSPT}
\end{align}

The green gauge generators become
\begin{subalign}
	\includegraphicsarray{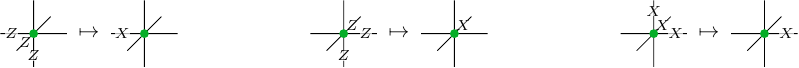} \\
	\includegraphicsarray{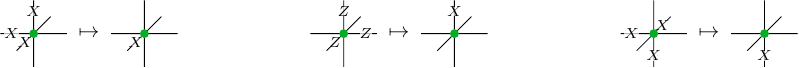}
\end{subalign}
while the yellow gauge generators become
\begin{subalign}
	\includegraphicsarray{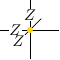}
	& \mapsto
	\includegraphicsarray{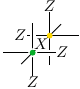}
	&         &
	\includegraphicsarray{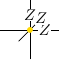}
	\mapsto
	\includegraphicsarray{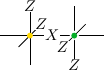}
	&         &
	\includegraphicsarray{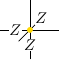}
	\mapsto
	\includegraphicsarray{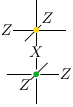}
	\\
	\includegraphicsarray{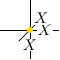}
	& \mapsto
	\includegraphicsarray{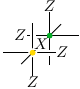}
	&         &
	\includegraphicsarray{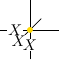}
	\mapsto
	\includegraphicsarray{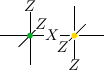}
	&         &
	\includegraphicsarray{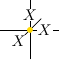}
	\mapsto
	\includegraphicsarray{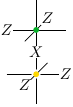}.
\end{subalign}
Since these all originate from the yellow gauge generators, they pairwise commute.

The most naive Hamiltonian is given by
\begin{align}
	\tilde{H}(J_g,J_y) = & -J_g\!\!\!\!\!\sum_{v,f\in\,\substack{\text{green} \\\text{spheres}}} (A_v+B_f)-J_y\!\!\!\!\!\sum_{v,f\in\,\substack{\text{yellow} \\\text{spheres}}} (A_v+B_f)
\end{align}
Following the mapping \cref{eqn:toSPT}, this becomes
\begin{align}
	H(J_g,J_y) & =-J_g\sum_{e\in\text{edges}} X_e -J_y\sum_{x\in\text{x-edges}} K_x -J_y\sum_{y\in\text{y-edges}} K_y-J_y\sum_{z\in\text{z-edges}} K_z, \label{eqn:SPTHamiltonian_app}
\end{align}
where
\begin{align}
	K_x:=
	\includegraphicsarray{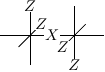}
	 &  &
	K_y:=
	\includegraphicsarray{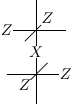}
	 &  &
	K_z:=
	\includegraphicsarray{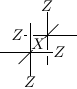}.
\end{align}

For all values $(J_g,J_y)$, this model is symmetric under the $\ZZ{2}\times\ZZ{2}$ 1-form symmetry generated by
\begin{align}
	\includegraphicsarray{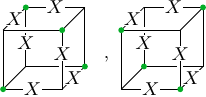},
\end{align}
in addition to (rigid) sheetlike symmetries along the $x-y$, $x-z$, and $y-z$ planes.
The model also has a global $\ZZ{2}$ symmetry $\otimes_e X_e$ for all parameters.

A self-duality map can be constructed by first applying the inverse of circuit \cref{eqn:toSPT} around all green vertices, then applying Hadamard to all edges, followed by applying \cref{eqn:toSPT} around all \emph{yellow} vertices.
Another (equivalent) duality can be constructed by removing all Hadamards from the previous.
Under these maps, the Hamiltonian transforms as
\begin{align}
	H(J_g,J_y)\mapsto H(J_y,J_g).
\end{align}
From this self-duality, we can deduce that if the model has a single phase transition, then it must occur where $J_g=J_y$.

\subsection{$J_y$ phase has SPT}

When $J_g\gg J_y$, the model \cref{eqn:SPTHamiltonian_app} is paramagnetic.
From the duality above, we find that when $(J_g,J_y)=(0,1)$, the ground state can be prepared from a paramagnet using a constant depth quantum circuit, precluding intrinsic topological order in the model.
The model has 1 qubit per edge, and one stabilizer per edge, so the ground state is unique.
This leaves open the possibility of symmetry-protected topological ordering (SPT).

The model \cref{eqn:SPTHamiltonian_app} is reminiscent of the Raussendorf-Bravyi-Harrington (RBH) model~\cite{Raussendorf2005}, although not obviously identical.
We demonstrate SPT at the $(0,1)$ point by identifying sheet-order parameters~\cite{Roberts2017} in \cref{sec:sheetOrder}, followed by gauging the symmetry in \cref{sec:gauging}.

\subsubsection{Sheet-order parameters}\label{sec:sheetOrder}
Consider an $x-z$ plane from the lattice, periodic at both boundaries.
The product of the following $K_x$ and $K_z$ in the plane leaves a sheet of $X$ operators on a subset of edges in the plane:
\begin{align}
	\includegraphicsarray{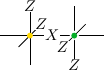}
	 &  &
	\includegraphicsarray{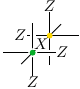}
	 &  &
	\includegraphicsarray{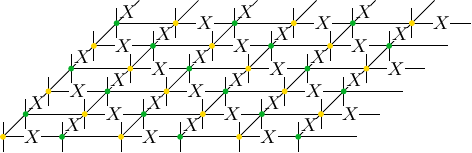}.\label{eqn:horizontalSheet}
\end{align}
Alternatively, the dual set of $K_x$, $K_z$ operators can be used to obtain the dual sheet.

Now, consider an $x-y$ plane, periodic at the $y$ boundary.
Restricting to a strip of the plane of size $~L_x/4$, the product of the following $K_x$ and $K_y$ operators gives a sheet with $X$ in the bulk, and $Z$ near the sheet boundary:
\begin{align}
	\includegraphicsarray{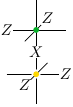}
	 &  &
	\includegraphicsarray{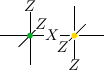}
	 &  &
	\includegraphicsarray{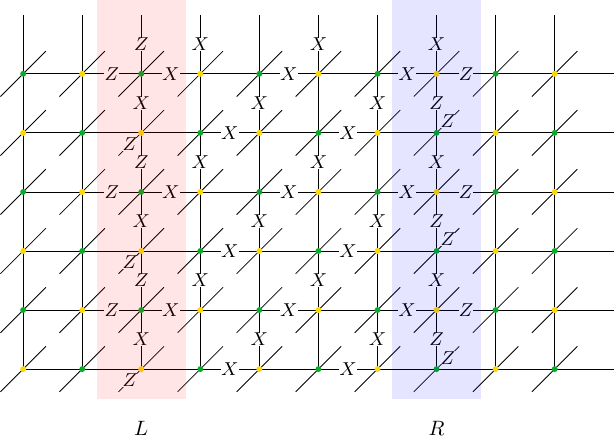}.\label{eqn:verticalSheet}
\end{align}
As above, the alternate set of $K_\bullet$ operators can be utilized to obtain the dual sheet.

Since these sheet operators are constructed from the stabilizers $K_\bullet$, they commute, and act as $+1$ on the ground space.
Conversely, in the trivial phase, the expectation value of the $x-y$ sheet is bounded to be below $1/2$~\cite{Roberts2017}.
This demonstrates that $H(0,J_y)$ is in a non-trivial SPT phase.

Since $H(0,1)$ is SPT non-trivial, $H(1,0)$ is SPT trivial, and $H(J_g,J_y)$ is symmetric for all parameter values, there must be a phase transition along the parameterized line $H(1,\lambda)$ for some $\lambda\in(0,1)$.
If this transition is unique, and continuous, then the naive Hamiltonian for the $\ZZ{2}$-model
\begin{align}
	\tilde{H}(J_g,J_y) = & -J_g\!\!\!\!\!\sum_{v,f\in\,\substack{\text{green} \\\text{spheres}}} (A_v+B_f)-J_y\!\!\!\!\!\sum_{v,f\in\,\substack{\text{yellow} \\\text{spheres}}} (A_v+B_f)
\end{align}
is gapless.
It could also be that the transition is first order, as in the symmetric sector~\cite{Li2023}, in which case the gap would remain open.

\subsubsection{Gauging the subsystem symmetry}\label{sec:gauging}
\begin{figure}\centering
	\includegraphicsarray{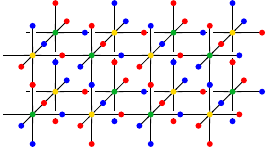}
	\caption{Given a fixed coordinate system, the edges of the lattice can be colored using the rule indicated here.}\label{fig:edgeColor}
\end{figure}

In addition to the sheet-order parameters considered above, we can \emph{gauge} the subsystem $\ZZ{2}\times\ZZ{2}$ symmetry.
It is convenient to color the edges as shown in \cref{fig:edgeColor}.
The gauging map $\Gamma$ is similar to that described in \onlinecite{Roberts2017,Yoshida2017}.
We refer to \onlinecite{Yoshida2017} for more details on gauging in general, and its application to identifying SPT phases.
States on the red/blue sublattice gauge to states on the blue/red sublattice.
\begin{align}
	\includegraphicsarray{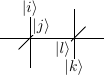}
	\overrightarrow{\Gamma}
	\includegraphicsarray{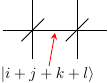}
	 &  &
	\includegraphicsarray{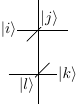}
	\overrightarrow{\Gamma}
	\includegraphicsarray{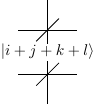}
	 &  &
	\includegraphicsarray{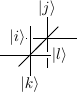}
	\overrightarrow{\Gamma}
	\includegraphicsarray{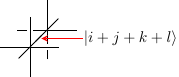}.
\end{align}

Applying this gauging map to the trivial Hamiltonian $H_0=-\sum_e X_e$ yields
\begin{align}
	H_{\Gamma} & =
	-\sum
	\includegraphicsarray{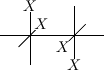}
	-\sum
	\includegraphicsarray{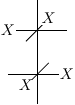}
	-\sum
	\includegraphicsarray{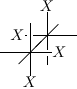}
	-\sum
	\includegraphicsarray{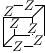},\label{eqn:trivialGaugedHamiltonian}
\end{align}
where the final terms act trivially on the gauge invariant subspace.
It is straightforward to check that this Hamiltonian contains no interactions between the two sublattices.
Examining the excitations of \cref{eqn:trivialGaugedHamiltonian}, we find that strings of $X$ on either sublattice create pairs of point-like excitations in the volumes at their endpoints, while membranes of $Z$ create loop-like excitations at their boundary.
This model is therefore topologically ordered, corresponding to two copies of the three-dimensional toric code model, albeit on an unusual lattice.

The Hamiltonian \cref{eqn:SPTHamiltonian_app} with $(J_g,J_y)=(0,1)$ is self dual under gauging, followed by transversal Hadamard operators on every edge.
Since this Hamiltonian can be obtained from the trivial paramagnet via a constant depth circuit, it cannot be topologically ordered.
Inequivalence of the gauged Hamiltonians implies that this Hamiltonian has SPT order.

\end{document}